\documentclass[structabstract]{aa}  
\usepackage{graphicx}
\usepackage{amsmath}
\usepackage[]{natbib}
\bibpunct{(}{)}{;}{a}{}{,} 
\usepackage{txfonts}
\begin{document}
   \title{A model for the non-thermal emission of the very massive colliding-wind binary HD~93129A} 

   \titlerunning{Non-thermal emission from HD~93129A}
   \authorrunning{S. del Palacio, V. Bosch-Ramon, G. E. Romero, P. Benaglia}


   \author{Santiago del Palacio
          \inst{1,2} \thanks{Fellow of CONICET}, Valent\'i Bosch-Ramon \inst{3}, 
          Gustavo E. Romero\inst{1,2} \thanks{Member of CONICET}
          \and Paula Benaglia\inst{1,2} $^{\star \star}$
          }

   \institute{Instituto Argentino de Radioastronom\'{\i}a (CCT La Plata, CONICET), C.C.5, (1894) Villa Elisa, 
   Buenos Aires, Argentina\\
                 \email{[sdelpalacio,romero]@iar-conicet.gov.ar}
                 \and
Facultad de Ciencias Astron\'omicas y Geof\'{\i}sicas, Universidad Nacional de La Plata, Paseo del Bosque, B1900FWA 
La Plata, Argentina
         \and
	 Departament d'Astronomia i Meteorologia, Institut de Ci\`ences del Cosmos (ICCUB),
Universitat de Barcelona (IEEC-UB), Mart\'{\i} i Franqu\`es 1, E-08028 Barcelona, Catalonia, Spain\\
             \email{vbosch@am.ub.es}
             }

   \date{Received ; accepted }

 
  \abstract
   {Recently, the colliding wind region of the binary stellar system HD~93129A was resolved for the first time using VLBI. This system, 
 one of the most massive known binaries in our Galaxy, presents non-thermal emission in the radio band, which can be used to infer
 the physical conditions in the system, and make predictions for the high-energy band.}
   {We intend to constrain some of the unknown parameters of HD~93129A through modelling the non-thermal emitter. We also aim at analysing
   the detectability of this source in hard X-rays and $\gamma$-rays.
   Finally, we want to predict how the non-thermal emission will evolve in the forthcoming years, 
   when the stars approach periastron.}
   {A broadband radiative model for the WCR has been developed taking into account the evolution of the accelerated particles     
 streaming along the shocked region, the emission by different radiative processes, and the attenuation of the emission
 propagating through the local matter and radiation fields. We reproduce the available radio data, and make predictions
 of the emission in hard X-rays and $\gamma$-rays under different assumptions.}
   {From the analysis of the radio emission, we find that the binary HD~93129A is more likely to have a low inclination 
   and a high eccentricity, with the more massive star being currently closer to the observer. The minimum energy of the non-thermal 
   electrons seems to be between $\sim 20 - 100$~MeV, depending on the intensity of the magnetic field in the wind-collision region.
   The latter can be in the range $\sim 20 - 1500$ mG.} 
   {Our model is able to reproduce the observed radio emission, and predicts that the non-thermal radiation from HD~93129A 
 will increase in the near future. With instruments such as \textit{NuSTAR}, \textit{Fermi}, and CTA, it will be possible 
 to constrain the relativistic particle content of the source, and other parameters such as the magnetic field strength in 
 the wind collision zone, which in turn can be used to obtain upper-limits of the magnetic field 
 on the surface of the very massive stars, thereby inferring whether magnetic field amplification is taking place in 
 the particle acceleration region.}
   \maketitle
%


\section{Introduction}\label{sec:intro}


Stars of type O and early B, and their evolved counterparts, Wolf-Rayet (WR) stars, are very massive and luminous.
These stars possess powerful stellar winds with mass-loss rates $\dot{M} \sim 10^{-7} - 10^{-5}$ M$_\odot$ yr$^{-1}$ and wind terminal 
velocities up to $v_\infty \sim 1000-3000$ km s$^{-1}$. If two massive stars, of types OB+OB or OB+WR, form a binary system, their 
winds will collide. These systems are known as massive colliding-wind binaries (CWBs). 
The stars in CWBs can be of various luminosity classes, from main-sequence to supergiants and WRs, 
and thus have different chemical abundances, mass-loss rates, and wind terminal velocities. The periods of CWBs 
also span a wide range. These systems, therefore, occupy a quite large region in the parameter space. In this work we focus
on the subgroup of CWBs capable of accelerating relativistic particles 
\citep[particle-accelerating colliding-wind binaries --PACWBs--;][]{2013A&A...558A..28D}.

The radio emission from PACWBs has three contributions: two thermal, namely one from the individual 
stellar winds and one from the thermal plasma in the wind-collision region (WCR); and the other one non-thermal (NT),
from the relativistic particles in the WCR. The latter is of synchrotron nature. An adequate interpretation of the radio spectral 
information requires the separation and analysis of both thermal and non-thermal components.
The steady thermal radio emission from a single stellar wind is due to the free-free process, and presents a flux density 
at a frequency $\nu$ of the form $S_\nu \propto \nu^\alpha$, with a spectral index $\alpha \sim 0.6$ \citep{1975MNRAS.170...41W}.
The thermal emission from the WCR is optically thick in radiative WCR shock systems, for which $\alpha \sim 2$, whereas it is optically thin in 
adiabatic WCR shock systems, leading to $\alpha \sim -0.1$. Such emission generally presents orbital modulation depending on the viewing angle 
\citep{Pittard2010II}.
In contrast, the features of NT radio emission are a spectral index much lower (often negative) than the typical 0.6 value for single stars, 
variable radiation and/or spectral index, and a flux density at relatively low frequencies much higher than the thermal component 
\citep{2000MNRAS.319.1005D}. In a few cases, VLBI observations have resolved the NT emission region associated with the 
WCR \citep[see, e.g.,][]{2005mshe.work...81O,2010RMxAC..38...41B}. 

Typically, the radio spectrum of a PACWB can be separated in three regions:
\begin{itemize}
 \item At $\nu > 10$ GHz, the thermal component of the winds is dominant. The flux density
 depends mostly on $\dot{M}$ and $v_\infty$. In principle, 
  constraints on such parameters can be obtained by the study of the spectrum at high frequencies.
  However, there are some caveats due to the uncertainties in the clumpiness of the winds 
\citep[e.g.][]{2011BSRSL..80...67B}. The thermal emission from the WCR can also contribute significantly to 
the total flux in compact binary systems.
 \item Approximately for $ 2 < \nu < 10$ GHz, the NT component is dominant. 
  The flux density is determined by the NT particle energetics and the $B$-field strength in the WCR. 
  No appreciable absorption features are expected unless the binary is rather compact.
 \item At $\nu < 2$ GHz, the absorption effects on the emission become important. Through the spectral 
 turn-over frequency and the steepness of the radiation below the cutoff, it is possible to infer whether 
 the suppression of the low energy emission is generated either by free-free absorption (FFA) in the stellar wind, 
 synchrotron-self absorption (SSA) within the emitter, attenuation due to the Razin-Tsytovich effect (R-T), or 
 because of a low-energy cutoff in the 
 electron energy distribution \citep{1980panp.book.....M}. If FFA is dominant, it is possible to constrain the 
 electron number density ($n_\mathrm{e}$) in the wind, and therefore the $\dot{M}$ of the star. If R-T dominates, 
 it is possible to constrain the value of $B/n_\mathrm{e}$ in the WCR. Finally, if there is a low-energy cutoff 
 in the electron distribution, it gives insight into the physics of particle acceleration processes in dense 
 and highly energetic environments. In these dense environments, unless the radio emitter is very compact, SSA 
 is likely to be negligible \citep[e.g.][]{2015MNRAS.451...59M}.
\end{itemize}

The synchrotron radio emission is explained by the presence of relativistic electrons and a local magnetic 
field \citep{Ginzburg1965, 2007A&ARv..14..171D}.
The detection of NT radio emission from many massive binaries, 
as well as hard X-rays \citep[see][for a detection in $\eta$-Carinae and a possible NT component in WR~140, 
respectively]{Leyder2010, Sugawara2015} and $\gamma$-rays (see below) in a few of them, 
leads to the conclusion that relativistic particle acceleration 
is taking place in these systems; electrons, protons and heavier nuclei can be efficiently accelerated to high 
energies (HE) through first-order diffusive shock 
acceleration (DSA) in the strong shocks produced by the collision of the supersonic stellar 
winds \citep{1993ApJ...402..271E, 2003A&A...399.1121B, 2006ApJ...644.1118R, 2006MNRAS.372..801P}. 

The data at radio-frequencies allow a characterisation of the injection spectrum of the relativistic
particles responsible for this emission, which in turn can be extrapolated to model the broadband spectral 
energy distribution (SED). The NT emission from IR to soft X-rays is completely
overcome by the thermal emission from the stars and/or the WCR, but beyond this range we can use the
information gathered at radio frequencies to predict the behaviour of the systems in the HE
domain, where the NT radiation dominates the spectrum again. The same population
of particles (electrons) producing synchrotron radiation also interacts with the stellar UV field,
producing HE photons through IC scattering \citep{1993ApJ...402..271E, 2003A&A...399.1121B, 2014ApJ...789...87R}.
Note that the changing physical conditions in the WCR along the orbit, together with the
geometrical dependence of some of the most relevant HE processes, lead to a periodic variability in the spectrum.
The relevant HE processes are anisotropic IC and $\gamma$-$\gamma$ absorption in the stellar radiation field. 
The role of proton-proton ($p$-$p$) interactions in radio emission may be also considered under some circumstances.

In this work, we develop a model that allows us to derive physical information of a binary stellar system and the NT 
emitting region using the available radio data. With this model, we can also determine under which conditions the 
system HD~93129A, one of the earliest, hottest, most massive and luminous binary stellar systems in the Galaxy, is most 
suitable for being detected at $\gamma$-rays. So far only $\eta$-Carinae \citep{2009ApJ...698L.142T,Farnier2011} and (presumably) 
WR~11 \citep{Pshirkov2016, Benaglia2016} are seen in $\gamma$-rays. In this paper we shall discuss whether 
HD~93129A is a promising candidate to join the family of $\gamma$-ray emitting, massive-star binaries.


\section{Model}\label{sec:model}


Stellar winds collide forming an interaction region bounded on either side by the reverse shocks of the winds. The shocked 
winds are separated by a surface called the contact discontinuity (CD), located where the ram pressure components perpendicular to this 
surface are in equilibrium: $p_{1\perp}=\rho_1v_{1\perp}^2 = p_{2\perp}=\rho_2v_{2\perp}^2$, where $\rho_{1,2}$ and $v_{1,2\perp}$ are the
density and velocity of the two winds at the CD. The stellar winds follow a $\beta$-velocity law in the radial direction, 
$v_\mathrm{w}(r) = v_\infty (1 - R /r )^\beta $, but here we have simply considered $v_\mathrm{w}(r) = v_\infty$ as the system we are studying 
is sufficiently extended. 
The azimuthal component of the wind velocity is also neglected.

A complete broadband radiative model of the WCR must take into account the (magneto)hydrodynamics of the wind interaction zone, 
the acceleration of relativistic particles, the emission due to these particles, and the possible attenuation of radiation through 
different processes. Some detailed models for the NT particle evolution \citep{2006ApJ...644.1118R,
2014ApJ...789...87R} and the NT emission \citep{2003A&A...409..217D, 2006A&A...446.1001P, 2006MNRAS.372..801P, 2014ApJ...782...96R}
have been developed in the past years. 

\begin{figure}
    \centering
    \includegraphics[width=0.36\textwidth, angle=0]{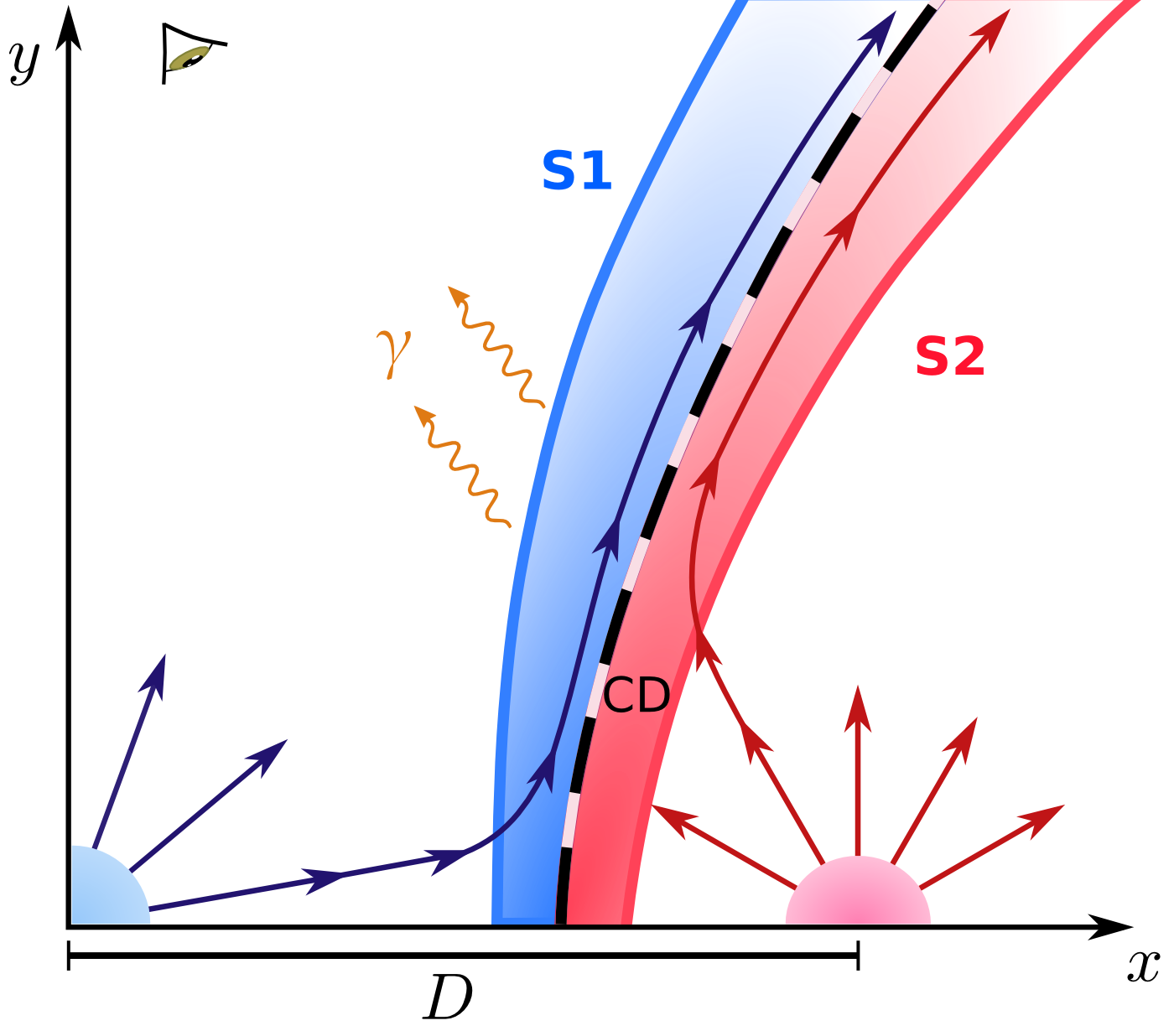}
    \caption[]{Sketch (not to scale) of the model considered in this work. The primary is located at $(0,0)$ and the secondary at $(D,0$). 
    The dashed line represents the contact discontinuity (CD), and on either side the shocked winds S1 and S2 are shown. The two solid lines 
    with arrows entering the WCR from both sides are representative of the different streamlines that conform the emitter.
    Photons produced in the WCR travel towards the observer through the unshocked stellar winds.}
    \label{fig:model}
  \end{figure}

One-zone models (i.e., punctual emitter) have been argued to be inadequate to represent CWBs because the free-free opacity 
varies along the extended region of synchrotron emission \citep{2003A&A...409..217D}. Thus, we 
develop a 2-dimensional (2D) model for the WCR approximating it as an axisymmetric shell with very small width, which 
effectively allows a 1D description for the emitter along which NT particles evolve. The shocked flows
stream frozen to the magnetic field, which is not dynamically dominant, starting at their injection/acceleration locations
along the WCR, from the apex to the periphery of the binary. A schematic picture is shown in Fig.~\ref{fig:model}. 
  
To compute the position of the CD, we use the Eq.~(6) prescribed in \cite{2004ApJ...611..434A}. As both shocks, S1 and S2,
have different properties because they depend on the conditions of the respective incoming wind \citep{2011A&A...530A..49B},
we apply an analytic hydrodynamical (HD) model to characterise the values of the relevant thermodynamical quantities 
for each side of the CD, which implies that there are actually two overlapped emitters at the CD. The WCR is formed by a sum of 
these 1D-emitters (linear-emitters hereafter) that
are symmetrically distributed around the two-star axis in a 3-D space: each discrete emission cell is first defined in the 
$XY$-plane, and the full 3-D structure of the wind interaction zone is then obtained via rotation around the $X$-axis. 
The hydrodynamics and particle distribution have azimuthal symmetry neglecting orbital effects. The dependence with the azimuthal
angle arises only for some emission and absorption processes that depend also on the position of the observer. 

A more detailed description of the source, the hydrodynamic aspects of the model, and the acceleration, emission, and absorption 
processes, is provided in the following subsections.

\subsection{Characteristics of HD~93129A}\label{sec:HD93129A}

The system HD~93129A is formed by an O2 If* star (the primary), and a likely O3.5 V star (the secondary). The primary is among the earliest, 
hottest, most massive, luminous, and with the highest mass-loss rate, O stars in the Galaxy. The astrometric analysis 
in \cite{Ben2015} indicates that the two stars in HD~93129A form a gravitationally bound system. Furthermore, the LBA data from 2008 
presented in \cite{Ben2015} revealed an extended arc-shaped NT source between the two stars, indicative of a WCR.

  \begin{table}
    \centering
    \begin{tabular}{lcc}
    \hline
    \textbf{Parameter} 		&	\textbf{Value}			&	\textbf{Unit}		\\
    \hline
    Primary spectral type	&	O2 If* $^{(a)}$			&			 	\\
    Secondary spectral type	&	O3.5 V $^{(b)}$			& 			 	\\
    System mass			&	$200 \pm 45$ $^{(c)}$		&	$M_\odot$ 		\\
    Distance			&	$d=2.3 \, ^{(d)}$		&	kpc			\\ 
    System separation (2008)	& $D_\mathrm{proj}= 36 \, ^{(e)}$	&	mas 		 	\\
    System bolometric luminosity & $L_\mathrm{bol}=6\times10^{39} \,^{(f)}$	&erg s$^{-1}$		\\
    Period			&	$P > 50 \, ^{(e)}$		&	yr		 	\\    
    Wind momentum ratio		&	$\eta=0.5 \, ^{(e)}$		&				\\
    
    \hline
   $T_{\mathrm{eff}_1}$		&	$42500 \, ^{(f)}$		&	K			\\
   $R_1$			&	18.3 $^{(g)}$			&	$R_\odot$		\\    
   $v_{\infty_1} \qquad$ ~(Primary)&	3200 $^{(h)}$			&	km s$^{-1}$		\\ 
   $\dot{M_1}$  		&	$10^{-5} \, ^\dagger$		&	$M_\odot$ yr$^{-1}$	\\ 
   $V_{\mathrm{rot}_1}/v_{\infty_1}$	&$\sim 0.1 \, ^{(i)}$		&	km s$^{-1}$		\\ 
   
    \hline
   $T_{\mathrm{eff}_2}$		&	44000 $^{(g)}$ 			&	K			\\
   $R_2$			&	13.1 $^{(g)}$			&	$R_\odot$		\\    
   $v_{\infty_2} \qquad$ ~(Secondary)&	$3000 \, ^{(j)}$		&	km s$^{-1}$		\\ 
   $\dot{M_2}$  		&	$5.3\times10^{-6} \, ^\dagger$	&	$M_\odot$ yr$^{-1}$	\\ 
   $V_{\mathrm{rot}_2}/v_{\infty_2}$	&$\sim 0.1 \, ^{(i)}$		&	km s$^{-1}$	 	\\ 

    \hline
    \end{tabular}
    \caption{Parameters of the system HD~93129A for the primary (sub-index 1) and the secondary (sub-index 2).
    $^\dagger$In this work we assumed a conservative value between those inferred from radio 
    \citep{Ben2015} and X-ray \citep{2011MNRAS.415.3354C} observations, consistent with a volume filling factor of the wind $f=0.07$ (see text).
    $^{(a)}$\cite{Walborn2002}; $^{(b)}$\cite{2006PASA...23...50B}; $^{(c)}$\cite{Maiz2008RMx}; $^{(d)}$\cite{2011ApJS..194....5G};
    $^{(e)}$\cite{Ben2015}; $^{(f)}$\cite{Repolust2004}; $^{(g)}$\cite{Muijres2012}; $^{(h)}$\cite{Taresch1997}; $^{(i)}$\cite{Mullan1984};
    $^{(j)}$\cite{Prinja1990}.}
    \label{table:parameters}
  \end{table}
  
\cite{2006PASA...23...50B} showed that the flux density of HD~93129A diminished with frequency, and pointed out a putative spectral 
turnover at $\nu < 1.4$~GHz, indicated by the slight decrease in the ratio between the 1.4~GHz and the 2.4~GHz flux.
Archival ATCA data show a point source with a change in flux level between 2003 and 2008. The emission is consistent 
with a NT spectral index $\alpha$ from $-1.2$ to $-1$, which corresponds to an injection index of $p = -2\alpha+1 \sim 3.2$. 
At frequencies above 15 GHz the thermal wind contribution dominates, as seen in Fig.~\ref{fig:dataradio}. The thermal emission, 
either from the stellar wind or the WCR, has positive spectral index \citep[e.g.][Fig.1, middle panel]{Pittard2010II}, and therefore 
a thermal contribution in the observed emission from HD~93129A would only imply an even steeper (i.e., more negative) intrinsic non-thermal spectral index.
For simplicity we assume here that the steep emission from HD~93129A is non-thermal in nature (which, given the hardness of thermal radiation,
is surely the case at the lowest frequencies).

Using the radio data, both the total fluxes and the wind momentum ratio, \cite{Ben2015} derived mass-loss rates in the range 
$2-5 \times 10^{-5}$ $M_\odot$~yr$^{-1}$, which are consistent with estimates obtained by other authors. However, these values might be
overestimated because of wind clumping. \cite{2011MNRAS.415.3354C} analysed \textit{Chandra} observations of HD~93129A 
and showed that the intrinsically hard emission from the WCR contributes less than 10\% to the total X-ray flux. 
The spectrum is dominated by thermal ($kT = 0.6$~keV) emission along with significant wind absorption. 
The broadband wind absorption and the line profiles provided two independent measurements of the wind mass-loss rate, 
$\dot{M} = (5.2 - 6.8) \times 10^{-6} M_\odot$~yr$^{-1}$, $\sim 3-10$ times smaller than the values derived from radio observations. 
The X-ray line-profile diagnostic of the mass-loss rate can be complementary to that in radio as it is not 
affected by small-scale clumping, as long as the individual clumps are optically thin to X-rays. Previous X-ray observations with 
\textit{XMM-Newton} also showed a thermal spectrum \citep{2006PASA...23...50B}. 

\cite{Ben2015} estimated some orbital parameters, although the binary has a very long period of more 
than 50 years \citep{2007ASPC..367..179B}, and thus the orbit is still poorly sampled and the fitting errors are
large. A high orbital inclination angle ($i \sim 77^\circ$) was favoured (nearly edge-on), with a relatively small eccentricity, although 
a highly eccentric orbit along with a small inclination angle (nearly face-on) could not be discarded. The 
angular separation between the components was $D_\mathrm{proj} = 55$ mas in 2003, and it shortened to 36 mas in 2008 
and 27 mas in 2013 \citep{2014ApJS..215...15S}. According to the estimates of the orbital parameters, the binary system should go through 
periastron in the next few years. An unpublished analysis by \cite{2007hst..prop11294M} also indicates that the system may be 
approaching periastron in $\sim 2020$ \citep{2011ApJS..194....5G}. In Tab.~\ref{table:parameters} the known system parameters, as well 
as others derived for the model described in Sect.~\ref{sec:model}, are listed. 
  
Previous estimates from \cite{2007ASPC..367..179B} indicated that the equipartition magnetic field at the WCR would be $\sim 20$~mG,
and the surface stellar magnetic field $B_* \sim 500$~G, whereas in \cite{2006PASA...23...50B} a smaller value of $B_* = 50$~G 
was derived assuming an initial toroidal component of the wind velocity of $v_\mathrm{rot} = 0.2 v_\infty$, an Alfv\'en 
radius of $r_\mathrm{A} = 3R_*$, and a stellar radius of $R_* = 20\,R_\odot$. The analysis presented in this work for two representative
cases of low and high stellar magnetic field gives consistent results (Sect.~\ref{sec:results}), with values of $B_*$ in the range $20-1000$ G.

For the two unshocked winds we assume the same molecular weight, $\mu_{\mathrm{w}1} = \mu_{\mathrm{w}2} = 1.3$, an 
rms ionic charge of $Z_{\mathrm{w}1} = Z_{\mathrm{w}2} = 1.0$, and an electron-to-ion ratio 
of $\gamma_{\mathrm{w}1} = \gamma_{\mathrm{w}2} = 1.0$ \citep{1995ApJ...450..289L}. For the WCR, the adopted stellar wind 
abundances are $X=0.705$, $Y=0.275$, and $Z=0.02$, considering complete ionization, which yields 
$\mu_1 = \mu_2 = 0.62$, $Z_1 = Z_2 = 1.23$, and $\gamma_1 = \gamma_2 = 1.34$. 

We are left with only a few free parameters: the value of $B$ in the WCR, the acceleration efficiency, the spectrum of 
relativistic particles, and the inclination of the orbit, whose fit has large uncertainty. As we will show in the following 
sections, it is possible 
to constrain $B$, the low-energy particle spectrum, and $i$, by using the radio fluxes measured for different epochs, and 
the morphological information from the resolved NT emission region. A list including the available radio data for different 
epochs and frequencies is presented in Tab.~3 of \cite{Ben2015}. 

\subsection{Hydrodynamics}

The hydrodynamics of the wind interaction in the adiabatic regime, and close enough to the binary to avoid orbital 
effects, is characterized to a large extent by the secondary-to-primary wind momentum
ratio $\eta=\dot{M}_2v_{\infty 2}/\dot{M}_1v_{\infty 1}$. If the wind shocks are adiabatic, the WCR thickness will be larger 
than in the case when the shocks are radiative, although for simplicity we assume that they are thin enough to neglect 
their width. Adiabatic wind shocks are convenient because the shocked two-wind 
structure is more stable, so a stationary approximation is more valid.
A cooling parameter $\Xi$ can be introduced as a characteristic measure of the importance of cooling in the shocked region. 
Using Eq.~(8) in \cite{1992ApJ...386..265S}, one obtains 
$\Xi = t_\mathrm{rad}/t_\mathrm{esc} \sim (v_8)^4 D_{12}/\dot{M}_{-7} \gtrsim 100 \gg 1$, 
and therefore one can consider that the shocked gas is adiabatic \citep[see, e.g.,][]{2011ApJ...743....7Z}.  
The value of $\Xi$ is computed assuming that $v_\mathrm{w}$ and $\dot{M}$ are similar for both stars, and that the 
shock is farther than 8~AU from any of the stars, roughly the distance at periastron.

Following the approach by \cite{2004ApJ...611..434A}, we assume that the flow along the WCR is laminar, neglecting any possible effect due to 
mixing of the fluid streamlines. We also suppose that the velocity of the fluid at a given position of the WCR
is equal to the projected tangential component of the stellar wind velocity. The particles are followed until they travel a distance $\sim 4D$, where 
$D$ is the linear separation between the stars. 
Notice that at large scales the Coriolis effect associated with the orbital motion breaks the azimuthal symmetry of the WCR, but this effect 
can be neglected for scales smaller 
than $\sim(v_\mathrm{w}/v_\mathrm{orb})\, D$, which is far enough even close to periastron for the emitting region considered.
  
  Each streamline coming from either star reaches the WCR at a different position (see Fig.~\ref{fig:model}). We consider a discrete number 
  of streamlines and that the $j$-streamline enters the WCR at the point $(x_j,y_j)$, where the $j$-th linear-emitter begins. For simplicity, 
  in our model the magnetohydrodynamic quantities at a cell $i$ are determined solely by the $j=i$ streamline and 
  are identical for all streamlines with $j < i$. To characterize the thermodynamical quantities in the WCR we consider the fluid to behave 
  like an ideal gas and we apply Rankine-Hugoniot jump conditions in the strong shocks.

\begin{figure}
    \centering
    \includegraphics[width=0.45\textwidth]{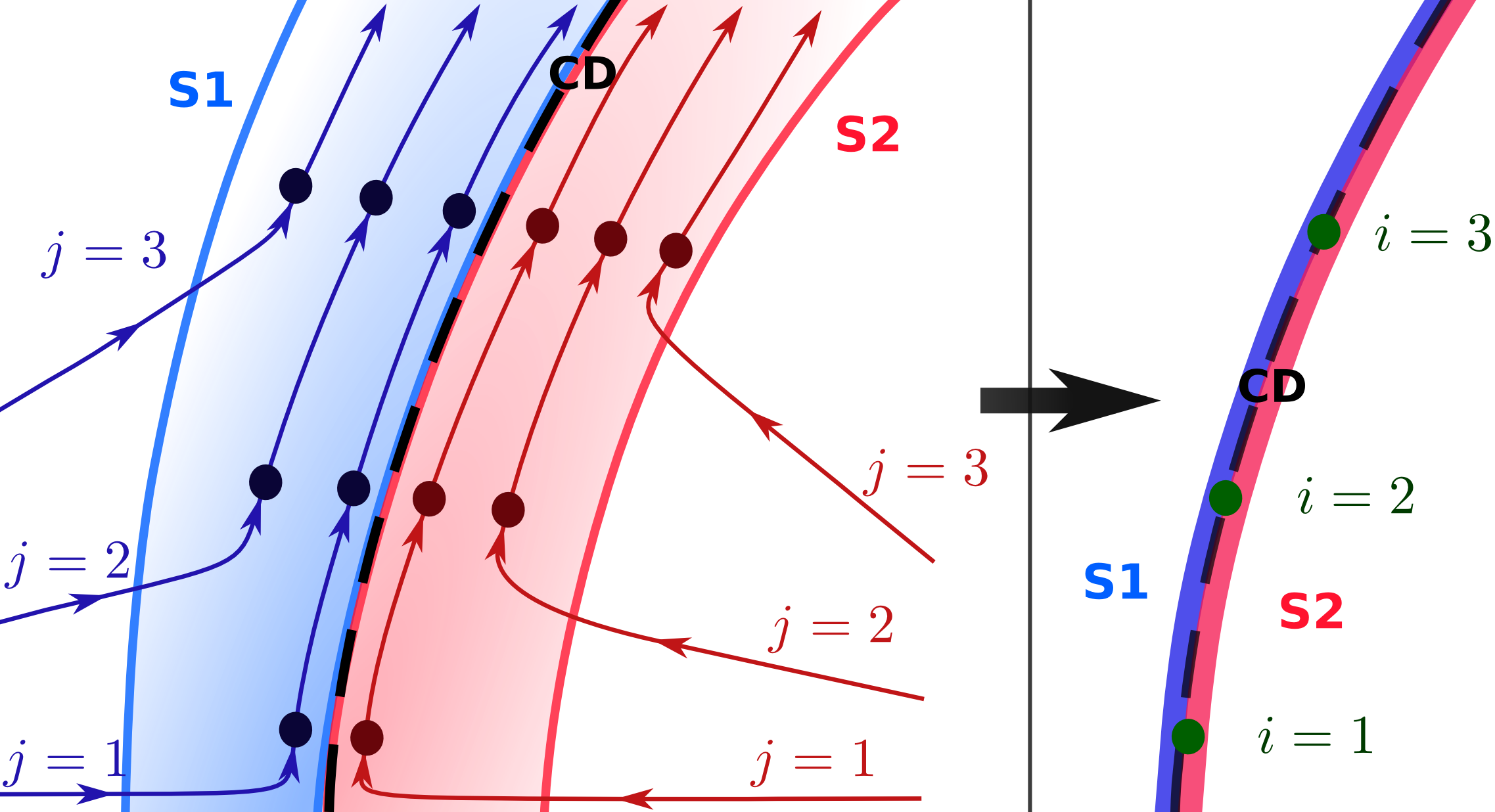} 
    \caption{Illustration of the model for the spatial distribution of NT particles in the WCR. On the \textit{left} side we show 
    different linear-emitters, named as $j=1,2,3$, coming from each star. On the \textit{right} side we represent a bunch of linear-emitters, 
    obtained from summing at each location the contributions from the different linear-emitters (see text).}
    \label{fig:lines}
\end{figure}

  Explicitly, the set of equations describing the fluid is the following:
    \begin{align}\label{eq:hydro}
    \rho(i) &= 4\rho_{\mathrm{wind}}(i) = \dot{M} \left[ \pi d(i)^2 v_{\mathrm{wind}} \right]^{-1}, \\
    P(i) &= 0.75\left( \rho_{\mathrm{wind}} v_{\mathrm{wind}\perp}^2 \right), \\
    v(i) &= \vec{v}_{\mathrm{wind}}(i) \cdot \vec{u}(i)_\parallel \,, \\
    T(i) &=  \frac{P(i) m_p \mu}{\rho(i)k}, \\
    B(i) &= \zeta^{1/2} B_{\mathrm{eq}} =  \left[\zeta \, 8\pi \gamma_\mathrm{ad} P(i) \right]^{1/2}.
    \end{align}
In those equations $i$ represents the position of the $i$-cell along the WCR (see Fig.~\ref{fig:lines}), 
$d$ is the distance to the respective star (i.e., to the primary for S1 and to the secondary for S2), 
$\gamma_\mathrm{ad} = 5/3$ is the adiabatic coefficient, $P$ is the pressure, $T$ is the temperature,
$v$ is the fluid velocity, $\vec{u}_\parallel$ is a versor tangent to the WCR (see Fig.~\ref{fig:lines}),
$m_p$ is the proton mass, $\mu$ is the mean molecular weight, and $\zeta$ the magnetic-to-thermal energy 
density ratio, $u_\mathrm{mag}/u_\mathrm{th}$ (with $u_\mathrm{th}=\gamma_\mathrm{ad} P$), at the shock. 
The parameter $\zeta$ is set to reproduce the observed radio fluxes, as described in Sect.~\ref{sec:results}. 
Finally, an upper limit for the stellar surface 
magnetic field can be estimated assuming that the magnetic field in the WCR comes solely from the adiabatic compression 
of the stellar field lines. Then, assuming an Alfv\'en radius $r_\mathrm{A} \sim R_*$, we get the expression 
$B_*=0.25 B(1) \left[ (D-x(1))/R_* \right](v_\infty/v_\mathrm{rot})$. We note that it is possible that the magnetic 
fields are generated or at least strongly amplified \textit{in situ}, and therefore the values of the upper-limits 
for the stellar surface magnetic field we derive could be well above the actual field values.

Numerical results of hydrodynamical simulations can also be used to obtain the conditions in which NT particles evolve and radiate 
\citep{2014ApJ...789...87R,2014ApJ...782...96R}. For simplicity, at this stage we use simple hydrodynamical prescriptions (Eqs.~1--5) 
because it eases the treatment. To first order, we do not expect significant differences from using more accurate 
fluid descriptions, as long as instabilities are not relevant in shaping the WCR. Nevertheless, in future works we plan
to use numerical results to characterize the hydrodynamics of the emitter.

\subsection{Particle distribution} \label{sec:distribution}

\begin{figure}
    \centering
    \includegraphics[width=0.3\textwidth, angle=270]{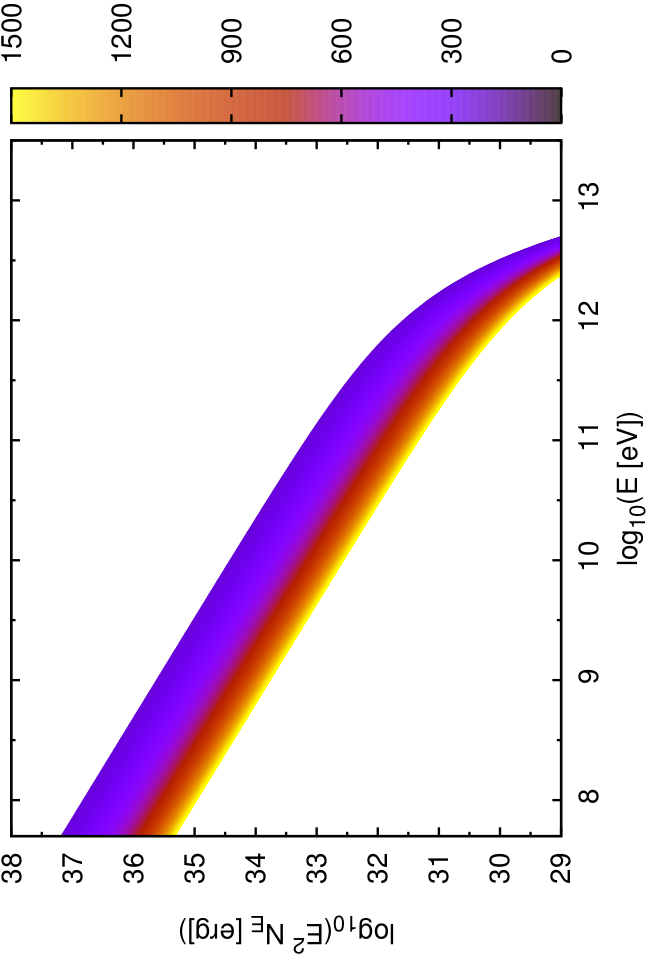} 
    \includegraphics[width=0.3\textwidth, angle=270]{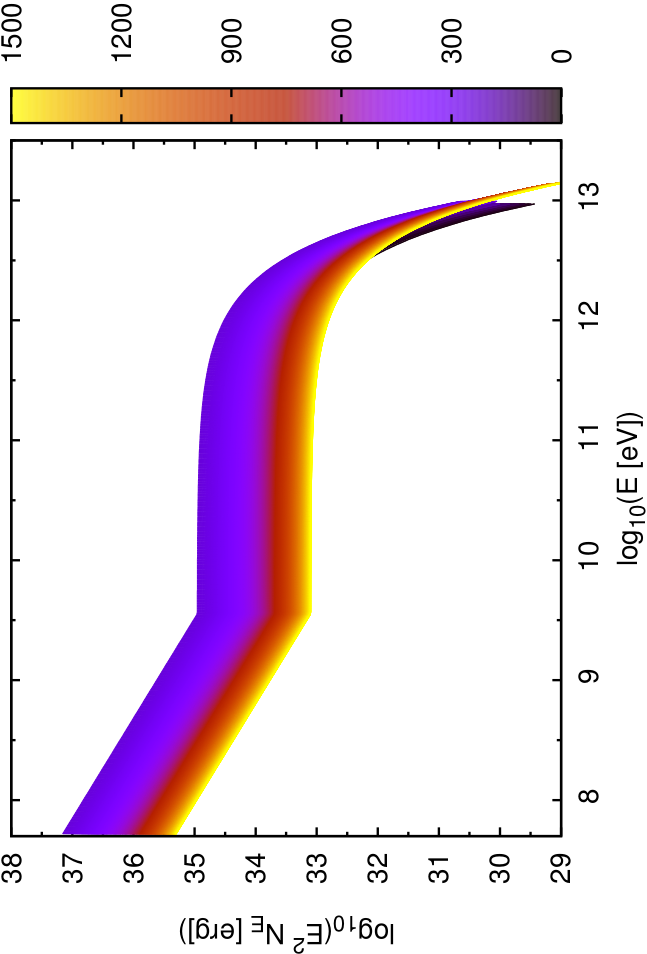}    
    \caption{Examples of the electron energy distribution for a bunch of linear-emitters assuming a constant spectral index (\textit{top}), 
    and a hardening at higher energies (\textit{bottom}). The colorbar corresponds to different positions along the line, with higher 
    values meaning farther from the apex. Both cases correspond to a low ambient magnetic field ($\zeta << 1$).}
    \label{fig:dist}
\end{figure}

The WCR presents hypersonic, non-relativistic adiabatic shocks, where NT particles of energy $E$ and charge $q_\mathrm{e}$
are expected to accelerate via DSA on a timescale 
$t_\mathrm{acc} \approx 2 \pi E (c/v_\infty)^2 \left( B c q_\mathrm{e} \right)^{-1}$~s \citep[see, e.g.,][for a review on 
non-relativistic DSA]{1983RPPh...46..973D}, where we have assumed diffusion in the Bohm regime. To derive 
the maximum particle energy, $t_\mathrm{acc}$ is compared with the relevant cooling and escape timescales. These particles are responsible of 
the observed NT radio emission. Radio observations indicate that the particle energy distribution of electrons that emit at 3-10 GHz has 
a power-law index of $p\sim 3.2$. The corresponding electron energy distribution seems particularly soft when compared with the observed 
spectral indices in other NT CWBs \citep{2013A&A...558A..28D}, and considering that the expected index for DSA in a strong shock is 
$p=2$. Such a soft energy distribution of the radio-emitting particles might be related to the wide nature of the orbit, which could imply
a stronger toroidal magnetic field component that may soften the particle distribution through quickly dragging the
particles away from the shock, at least at low energies. At this stage, we just 
take the radio spectrum as a constraint for the radio-emitting particles, and we do not discuss the details of their acceleration. The
particles accelerated in each cell of the WCR cool through various processes, although -except for the most energetic particles- advection 
dominates the particle energy loss. Under such condition the evolved particle distribution has the same spectral index as the 
injected distribution, as seen in Fig.~\ref{fig:dist}, because cooling cannot modify the distribution of the particles as they move along the 
WCR (see Eq.~\ref{eqnei} below). The energy distribution of the accelerated particles at injection is taken as $Q(E)\propto E^{-p}$.

Radio observations do not provide any information of the electron distribution above the corresponding electron energies. On the other hand,
the DSA process for the less energetic particles could be affected by small-scale irregularities in the shock precursor, magnetic field 
geometry (already mentioned), \textit{energy-dependent} compression ratios, non-linear effects in the shocks, 
etc. \citep[see, e.g.,][]{1983RPPh...46..973D}. 
In addition, other acceleration processes might also act besides DSA, such as magnetic reconnection \citep{2012MNRAS.423.1562F}, 
affecting as well the particle distribution. Therefore, the resulting particle distribution could be softer at low 
energies \citep[see, e.g.,][for similar conclusions in SN 1006]{2009A&A...505..169B}, and harder at high energies. To explore 
the possibility of a harder distribution of the accelerated particles at high energies, two different cases are investigated 
here (see Sects.~\ref{subsec:lowB} and \ref{subsec:harden}); in Fig.~\ref{fig:dist}, the two different models of the 
electron energy distribution are shown. 

Radio observations also show that radio emission is strongly reduced below 1-2 GHz. Our analysis below (Sect.~\ref{sec:results}) shows 
that statistically this reduction is best explained in terms of a low-energy cutoff in the electron energy distribution. The determination of 
the origin of such a low-energy cutoff is out of the scope of this work, but this may be again related to the peculiarities of particle
acceleration in HD~93129A. 
An electron with Lorentz factor $\gamma$, in presence of a magnetic field of intensity $B$, 
emits mostly at a frequency $0.29 \nu_\mathrm{c} \sim 10^6 \,B\, \gamma^2$. Therefore, to obtain a cutoff around $\nu \sim 1$~GHz one needs 
$B\, \gamma_\mathrm{min}^2 \sim 10^3$. For $B_\mathrm{WCR} \sim 1$ G this leads to $\gamma_\mathrm{min} \sim 30$, whereas for 
$B_\mathrm{WCR} \sim 10$ mG it leads to $\gamma_\mathrm{min} \sim 280$.

The normalization of the evolved particle distribution depends on the amount of kinetic energy flux perpendicular to the shock surface, 
and on the fraction of that flux that goes to NT particles ($f_\mathrm{NT}$). The luminosity associated with such a flux is only $1-6$\% 
of the total wind luminosity, and the explored range of $f_\mathrm{NT}$ is $\sim 0.01-100\%$, which yields a NT  
luminosity $L_\mathrm{NT} \sim 10^{32}-10^{36}$ erg s$^{-1}$. We notice that secondary particles produced in \textit{p-p} interactions 
\citep[e.g.,][]{Ohm2015} are not expected to play a dominant role; a brief analysis of this scenario is discussed in Sect.~\ref{subsec:hadronic}.

 \subsection{Numerical treatment}
 We apply the following scheme for each of the two shocks:
 \begin{enumerate}
  \item We integrate numerically the Eq.~(6) from \cite{2004ApJ...611..434A} to obtain the location of the CD on the $XY$-plane in discrete 
  points $(x_i,y_i)$. We characterize the position of the particles in their trajectories along the WCR through 1D cells located at those points. 
  
  \item We compute the thermodynamical variables at each position in the trajectory from Eqs.~(1)--(5). Those quantities at a point 
  $(x_i,y_i)$ only depend on the position characterized by the value of $i$ alone. 
  
  \item The wind magnetic field lines reach the wind shocks at different locations, from where they are advected along the WCR. 
  We simulate the different trajectories by taking different values of 
  $i_\mathrm{min}$: the case in which $i_\mathrm{min} = 1$ corresponds to a line starting in the apex of the WCR, whereas $i_\mathrm{min} = 2$ 
  corresponds to a line which starts a bit farther in the WCR, etc. The axisymmetry allows us to compute the trajectories only for the 
  1D emitters with $y > 0$.  
  
  \item Relativistic particles are injected in only one cell per linear-emitter (the corresponding $i_\mathrm{min}$), and each 
  linear-emitter is independent from the rest. For 
  a given linear-emitter, particles are injected at a location $\left(x_{i_\mathrm{min}},y_{i_\mathrm{min}}\right)$, where we estimate the 
  particle maximum energy $E_{0_\mathrm{max}}$ and the cooling $\dot{E_0}$. Adiabatic cooling is included as $t_\mathrm{adi} \approx R(i)/v(i)$, 
  where $R(i)$ is the distance from the $i$-cell to the closest star. In the Bohm regime, the diffusion coefficient is 
  $D_\mathrm{Bohm} = r_\mathrm{g}c/3$, where $r_\mathrm{g} = E/(qB)$ is the gyro-radius of the particle. 
  The diffusion timescale $t_\mathrm{dif} \approx 0.5 R(i)^2/D_\mathrm{Bohm}$
  is much larger than the radiative cooling timescale, and therefore diffusion losses are negligible. We calculate the advection time, i.e., 
  the time taken by the particles to go from one cell to the following one, as $t_\mathrm{adv} = d(i)/v(i)$, where 
  $d(i) = \sqrt{ (x_i-x_{i-1})^2 + (y_i-y_{i-1})^2 }$ is the size of that cell. We obtain the distribution at the injection point, 
  $N_0(E,i_\mathrm{min})$, as $N_0(E,i_\mathrm{min}) = Q(E) \, t_\mathrm{adv}$.
  
  In the next step in the trajectory of the HE particles, which are attached to the flow through the chaotic $B$-component, we 
  calculate $E_\mathrm{max}(i)$ and $\dot{E}(i)$ to obtain the version of the injected distribution evolved under the relevant energy 
  losses (IC scattering, synchrotron, and relativistic Bremsstrahlung for relativistic electrons; proton-proton interactions,
  adiabatic losses, and diffusion for relativistic protons):
  \begin{equation}
   N(E,i) = N_0(E,i_\mathrm{min})\frac{\dot{E_0}}{\dot{E}} \, \theta \left( E_\mathrm{max}(i)-E \right),
   \label{eqnei}
  \end{equation}
  where $\theta$ is the Heaviside function and $E_\mathrm{max}$ is the evolved maximum energy of the particles streaming away 
  from the injection point with $i=i_\mathrm{min}$. The particle energy distribution in the injection cell is normalised according to the 
  NT luminosity deposited in that cell. The normalization depends both on the size of the cell, as lines are not equi-spatially
  divided, and on the angle between the direction perpendicular to the CD and the unshocked wind motion, which varies along the CD. 
  
  \item We repeat the same procedure varying $i_\mathrm{min}$, which represents different line emitters in the $XY$-plane.
  
  \item Finally we calculate $N_\mathrm{tot}(E,i)$ summing the distributions $N(E,i)$ obtained 
  before. The number of sums depends on the value of $i$: for $i=1$ there is only one population; for $i=2$ we have the
  evolved version of the particles injected at $i_\mathrm{min}=1$, and those injected at $i_\mathrm{min}=2$ 
  (two streamlines total); 
  for $i=3$ there is the evolved version of the particles injected at $i_\mathrm{min}=1$ and $i_\mathrm{min}=2$ (two streamlines), and those 
  injected at $i_\mathrm{min}=3$ (giving a total of three streamlines crossing the point $i=3$), 
  and so on. Recall that along the CD there are actually two overlapping bunches of linear-emitters that correspond to S1 and S2. 
  An illustration of the linear-emitters and the bunches of linear-emitters is shown in Fig.~\ref{fig:lines}.
  \end{enumerate}

\subsection{Emission and absorption}

Once the particle distribution for each cell $(x_i,y_i)$ is known, it is possible to calculate the total emission corrected for absorption
as follows: 
\begin{enumerate}
 \item We rotate the bunch of linear-emitters around the axis joining the two stars, ending up with many bunches of linear-emitters, 
 each with a different value of the azimuthal angle $\phi$. The particle energy distribution is normalised accounting for the number 
 of bunches of linear-emitters with different $\phi$-values (i.e., for $m_\phi$ bunches of linear-emitters, each one has a luminosity
 $L_\mathrm{NT}/m_\phi$, as we are assuming azimuthal symmetry). In this way, we have the evolved particle energy distribution for 
 each position $(x_i,y_i,z_i)$ of the CD surface in a 3D geometry.
 
 \item At each position $(x_i,y_i,z_i)$, we calculate the total opacity $\tau$ in the direction of the observer 
 for $\gamma$-$\gamma$ \citep[e.g.,][]{Dubus2006} and free-free \citep[e.g.,][]{rybicki_radiative_1986} absorption. Note that 
 the FFA coefficient in the limit $h\nu << kT$ is $\alpha_\mathrm{ff} \propto n_\mathrm{e} n_\mathrm{i} T^{-3/2}$, where 
 $n_\mathrm{e}$ and $n_\mathrm{i}$ are the electron and ion number density, respectively. From Eqs.~(1)--(4) it follows that the 
 free-free opacity is much larger in the wind than in the WCR \citep[see Fig.~5 from][for numerical estimates of this effect]{2006MNRAS.372..801P}. 
 Thus, our consideration of a thin shock is not an impairment when calculating the impact of FFA.
 
 \item At each position $(x_i,y_i,z_i)$ we calculate the emission of the various relevant radiative processes, IC, synchrotron,
 \textit{p-p}, and relativistic Bremsstrahlung \citep[see, e.g.,][and references therein]{Bosch-Ramon2009}, and then take into 
 account the absorption when necessary. Except for the IC, which 
 depends on the star-emitter-observer geometry, the other radiative processes can be regarded as isotropic. This is related to the 
 fact that the magnetic field is expected to be highly tangled in the shocked gas. In those cases, i.e. excepting IC, the only
 dependence with the azimuthal angle appears through absorption effects.
 
 \item The total emission from each location for both S1 and S2, accounting for absorption if needed, is summed to obtain the SED of the source.
\end{enumerate}

 
\section{Results} \label{sec:results}


The particle energy distribution, and the NT emission through IC, synchrotron, and relativistic Bremsstrahlung are computed for 
different scenarios and epochs. The \textit{p-p} interactions are also discussed to assess their potential importance in HD~93129A.

\subsection{Inclination of the orbit} \label{subsec:incli}

We revisit a previous estimation of the inclination of the orbit ($i$) of HD~93129A obtained by \cite{Ben2015} based on the reconstruction 
of the observed orbit. In this work, we explore the different 
possibilities for $i$ under two constraints: the first is the total flux intensity, and the second is the morphology of the emission map.
As for the first, if $i$ is high (say, $i > 60^\circ$), it is hard to explain the low-energy cutoff in the SED by means of FFA in 
the stellar wind (see the forthcoming Sect.~\ref{sec:maps}). This is reflected by the high value ($>10$) of the reduced $\chi^2$ (3 dof) of the 
fit at radio frequencies. 
In principle it could still be possible to have $\chi^2 < 10$, reproducing better the low-energy cutoff, by assuming that the stellar wind is much denser, 
but such a scenario would be 
in tension with the observational X-ray data. The only possibility remaining is to assume a low-energy cutoff in the particle distribution, 
which yields $\chi^2 \gtrsim 1$. This leads us to the second point: we explored various values of $i$ and the subsequent emission maps
(see Sect.~\ref{sec:maps} below). In Fig.~\ref{fig:maps2GHz_variousi} we show that a high $i$ ($>60^\circ$) leads to synthetic maps inconsistent 
with the source radio morphology. We conclude that scenarios with a low $i$ are, therefore, strongly favoured, and we 
restrict our further analysis to them. For values of $i \sim 15^\circ$, the linear separation of the stars at the epoch of observation is 
$D \approx 85$ UA.

We note that a small improvement in the fit is obtained if the 
secondary is in front of the system, as its wind is slightly slower and therefore its column density is slightly higher, 
thereby increasing a bit the role of FFA in the generation of the low-energy cutoff seen in radio. In all the remaining calculations 
(including the maps presented in Fig.~\ref{fig:maps2GHz_variousi}) we consider the secondary to be closer to the observer. However, 
as the winds of both stars have similar characteristics and this effect is small, this should not be taken as a tight restriction. 

\begin{figure*}
    \centering
    \includegraphics[width=0.67\textwidth, angle=0]{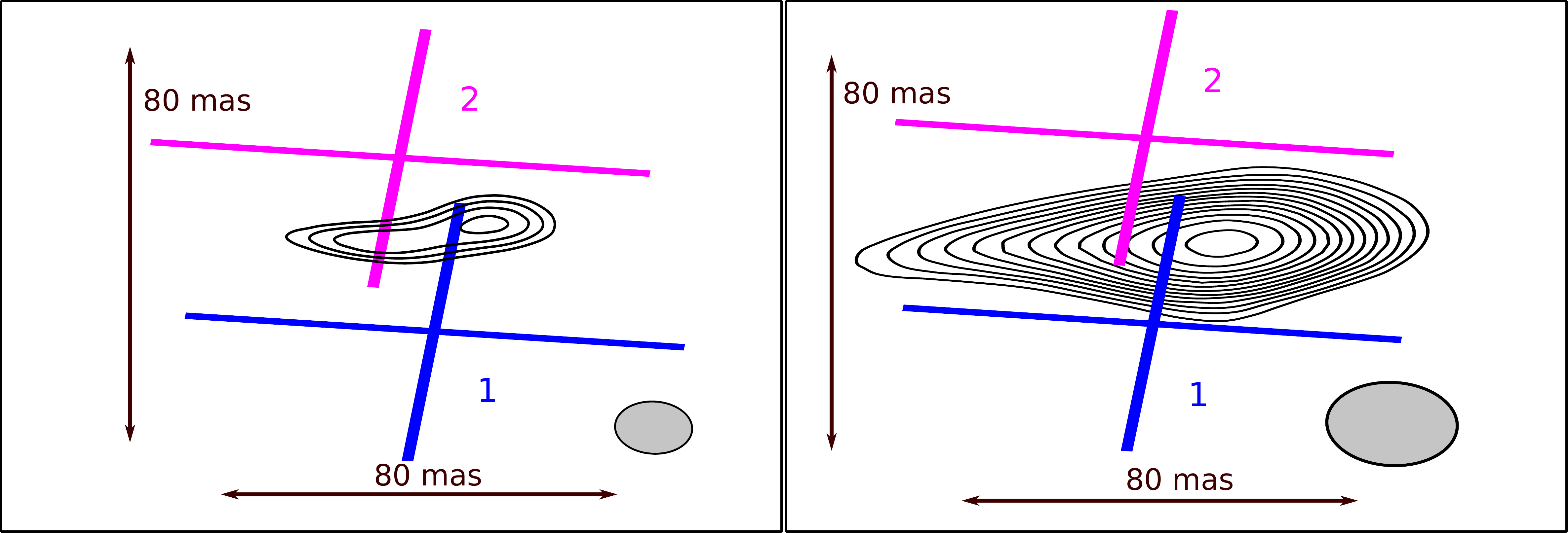}
    \includegraphics[width=0.47\textwidth, angle=270]{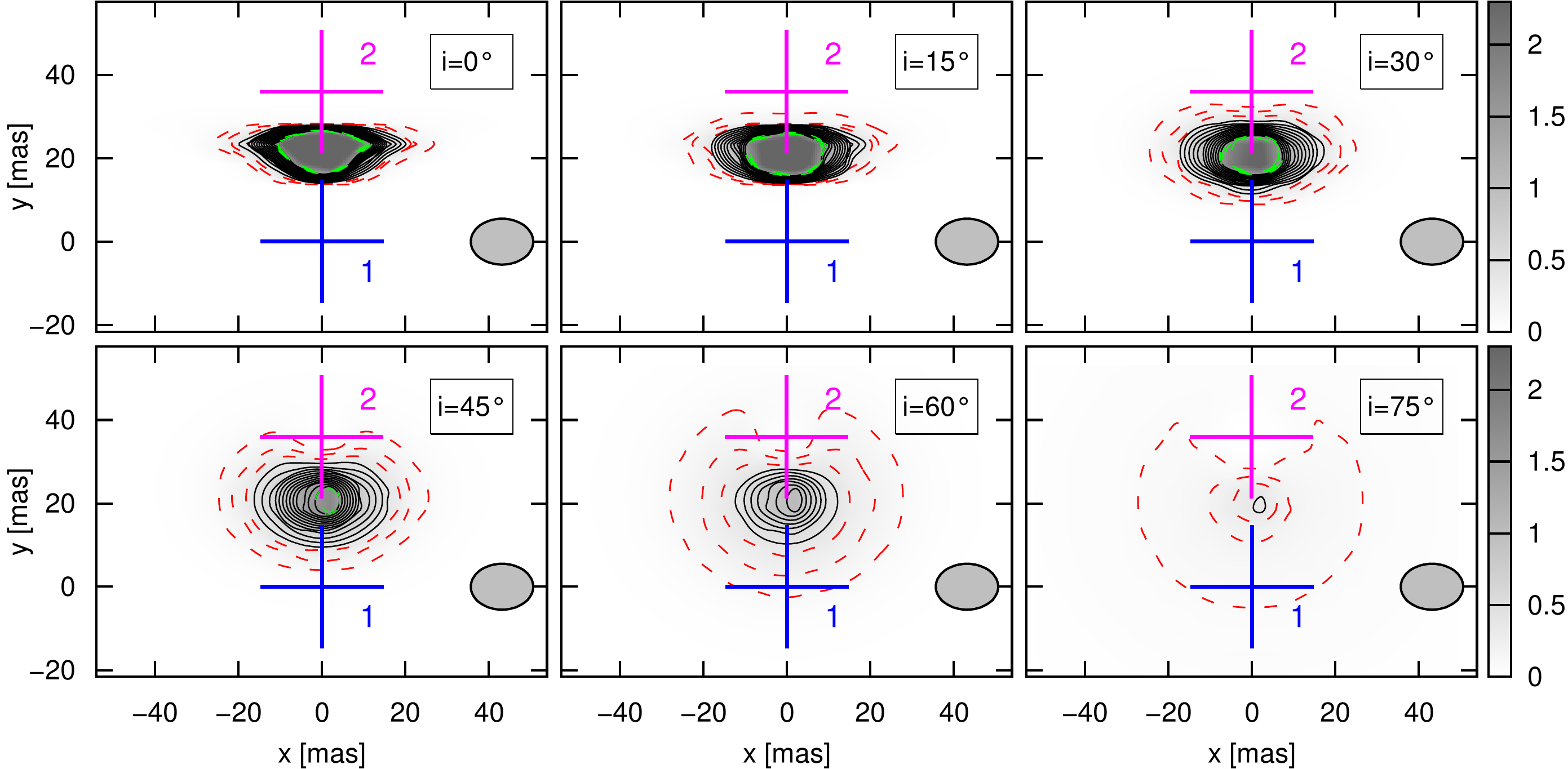}
    \caption[]{Comparisson between the observed radio emission map at 2.3 GHz (\textit{top}, adapted from \citealt{Ben2015}) 
    and our synthetic maps for the same binary separation as in the epoch of LBA observations. The two observed maps have different synthesized
    beams: the one on the \textit{left} has a better resolution but a poorer sensitivity, whereas the one on the \textit{right} has a poorer resolution
    but a higher sensitivity.
    In the synthetic maps, different values of the inclination of the orbit, ranging from $0^\circ$ to $75^\circ$, are considered. The colorbar 
    represents the flux in units of mJy beam$^{-1}$. The contours from the observed map start at
    0.4 mJy beam$^{-1}$ and increase in steps of 0.1 mJy beam$^{-1}$, with a maximum value of 1.6 mJy beam$^{-1}$ (\textit{right}). 
    In the synthetic maps, 
    along with these contours, we show the contours at 0.1, 0.2 and 0.3 mJy beam$^{-1}$ (below the 3$\sigma$ threshold of the observations) 
    in red, and the contours at 1.8, 2.0 and 2.2 mJy beam$^{-1}$ (above the observed values) in green. Cases where $i>60^\circ$ completely fail to 
    reproduce the observed maps. The best fit is achieved for low inclinations ($i < 30^\circ$).}
    \label{fig:maps2GHz_variousi}
\end{figure*}

\subsection{Low magnetic field scenario} \label{subsec:lowB}

We apply the model described in Sect.~\ref{sec:model} for a scenario with a low magnetic field (i.e., $\zeta << 1$). 
To reproduce the 2008 radio observations, we fix the values of the following parameters: $i=15^\circ$, motivated by the discussion in 
Sec.~\ref{subsec:incli}; $f_\mathrm{NT} = 0.11$, which represents an optimistic case, yet more conservative than assuming 
$f_\mathrm{NT}=1$; $\zeta = 2.\times10^{-4}$, which leads to $B_\mathrm{WCR}\sim 24$ mG, and $B_* < 30$ G; and electron minimum energy given by 
$\gamma_\mathrm{min} = 100$, which is needed to obtain the low-energy cutoff in the synchrotron spectrum (recall that this value 
is constrained by the value of $B_\mathrm{WCR}$ as described in Sec.~\ref{sec:distribution}). We also consider equipartition of energy between 
electrons and protons. Note that, since it is statistically favoured, the case with a high $E_\mathrm{min}$ is adopted even when FFA is
important for low $i$ values. A constant value of $p=3.2$, the one suitable for fitting the radio data, is adopted for the particle energy 
distribution at injection, $Q(E)$.

\begin{figure}
    \centering
    \includegraphics[width=0.3\textwidth, angle=270]{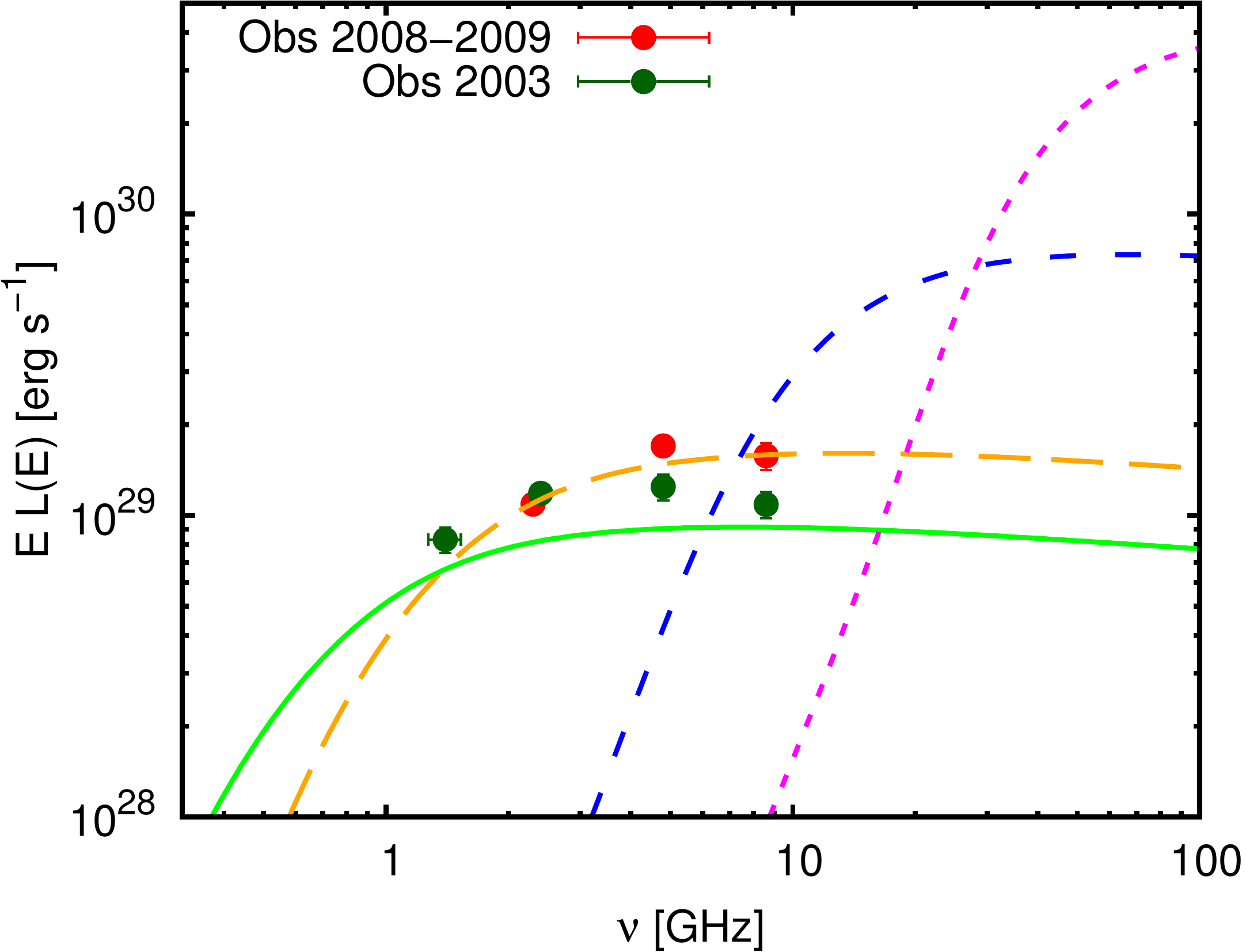}
    \caption{The curves represent the modeled emission for a distance between the componentes of $1.5\,D$ (green solid line), $D$ (orange 
    long-dashed line), $D/3$ (blue intermediate-dashed), and $D/10$ (magenta short-dashed). The goodness of the fit is given by 
    $\chi^2 \sim 17$ (3 dof) for the 2003-2004 observations, and $\chi^2 \sim 1.9$ (4 dof) for the 2008-2009 observations.}
    \label{fig:dataradio}
\end{figure}

We calculate the radio emission of the system for various epochs. First, when the distance between the two stars was $D$ (roughly the 
epoch of the 2008-2009 observations), achieving a good fit with a reduced $\chi^2 = 1.9$ (3 dof). To check the consistency 
in time of our results, we calculate the expected radio emission for the previous epoch of observation, 2003-2004, when the distance between 
the components was $1.5\,D$; we achieve a poor fit with a reduced $\chi^2=17$ (4 dof), 
which shows that an error in the fluxes $\sim 25\%$ arises from assuming that $\zeta$, $\gamma_\mathrm{min}$, $\alpha$ and 
$f_\mathrm{NT}$ remain constant in time. Finally, we repeat 
the calculations for closer distances, $D/3$ (roughly the present time), and $D/10$ (roughly the periastron passage). 
The slight departure of our predictions from the 2003--2004 observational points imply that the constant parameter assumption 
does not strictly apply, although the qualitative conclusions we arrive should be rather robust. 
In Fig.~\ref{fig:dataradio} we present the time evolution of the radio emission. The flux at 2.3~GHz is getting fainter with time 
due to increased FFA, as photons have to travel through denser regions of the winds. This effect is not so important at 8.6~GHz but 
for $D/10$, so the flux at 8.6~GHz increases until $D\sim D/3$, and then drops considerably for $D\sim D/10$. We conclude that in the near 
future (i.e., close to periastron), the low-frequency turnover in the spectrum will be due to FFA. 
Comparing past and future observations can show whether the evolution
of the low-frequency radio spectrum is consistent with two different
hardening mechanisms (low-energy cutoff in the electron distribution and FFA), or just one (FFA).
For such purpose, sensitive observations at frequencies below 1 GHz are recommended, as the steepness of the turnover will clarify 
the nature of the supression mechanism affecting the radio emission (being steeper if FFA dominates).
 
\begin{figure}
    \centering
    \includegraphics[width=0.275\textwidth, angle=270]{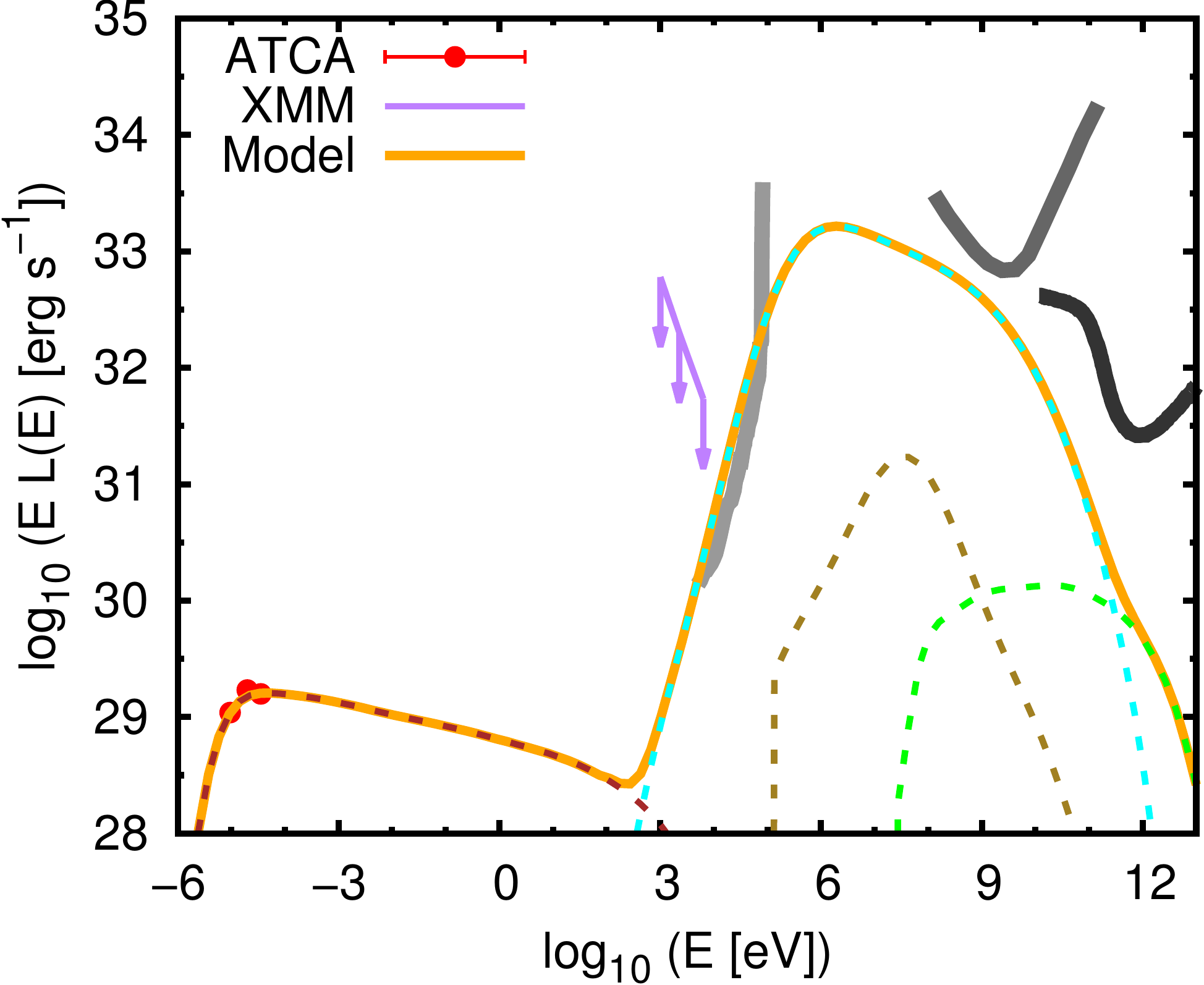}
    \includegraphics[width=0.275\textwidth, angle=270]{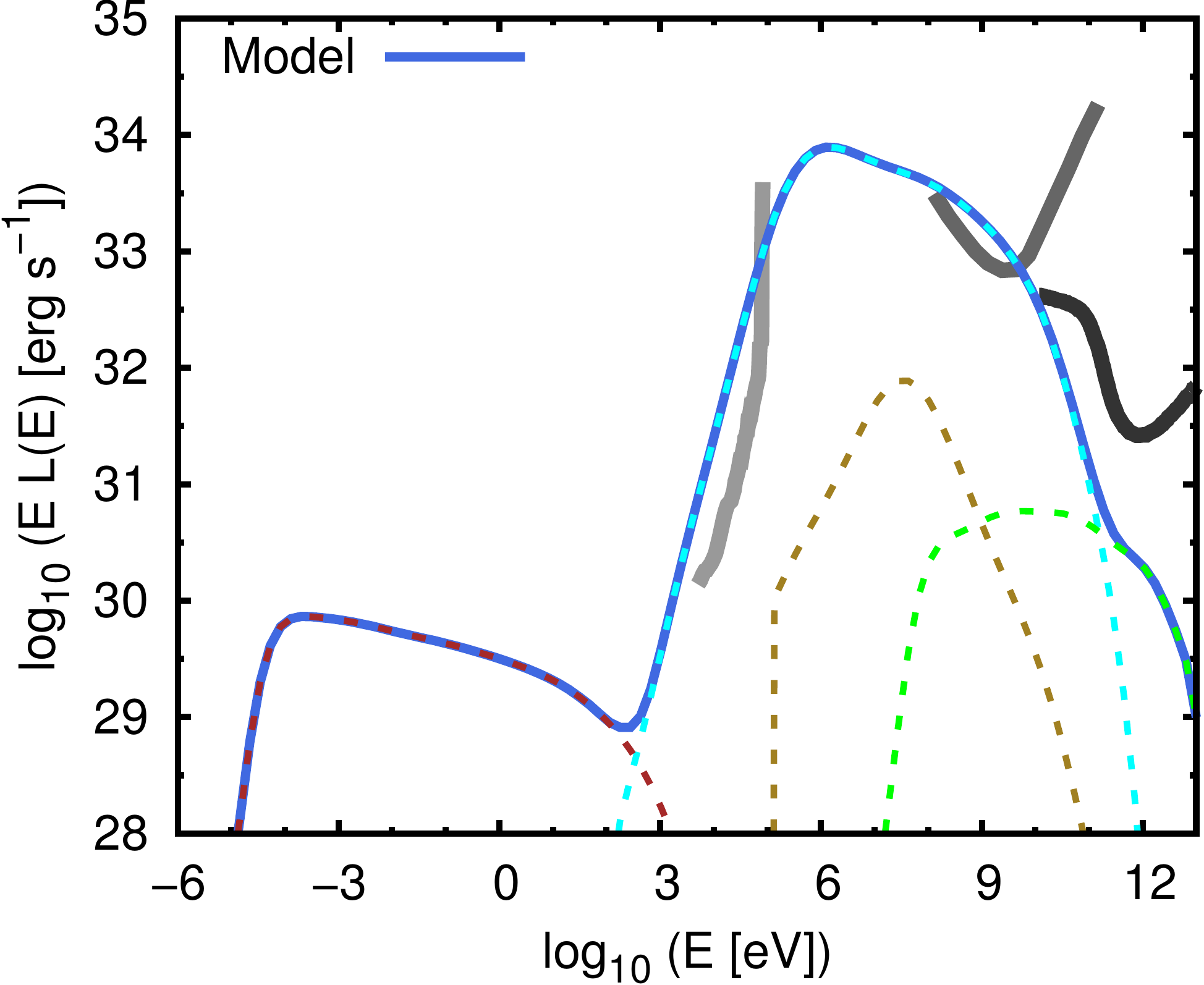}
    \includegraphics[width=0.275\textwidth, angle=270]{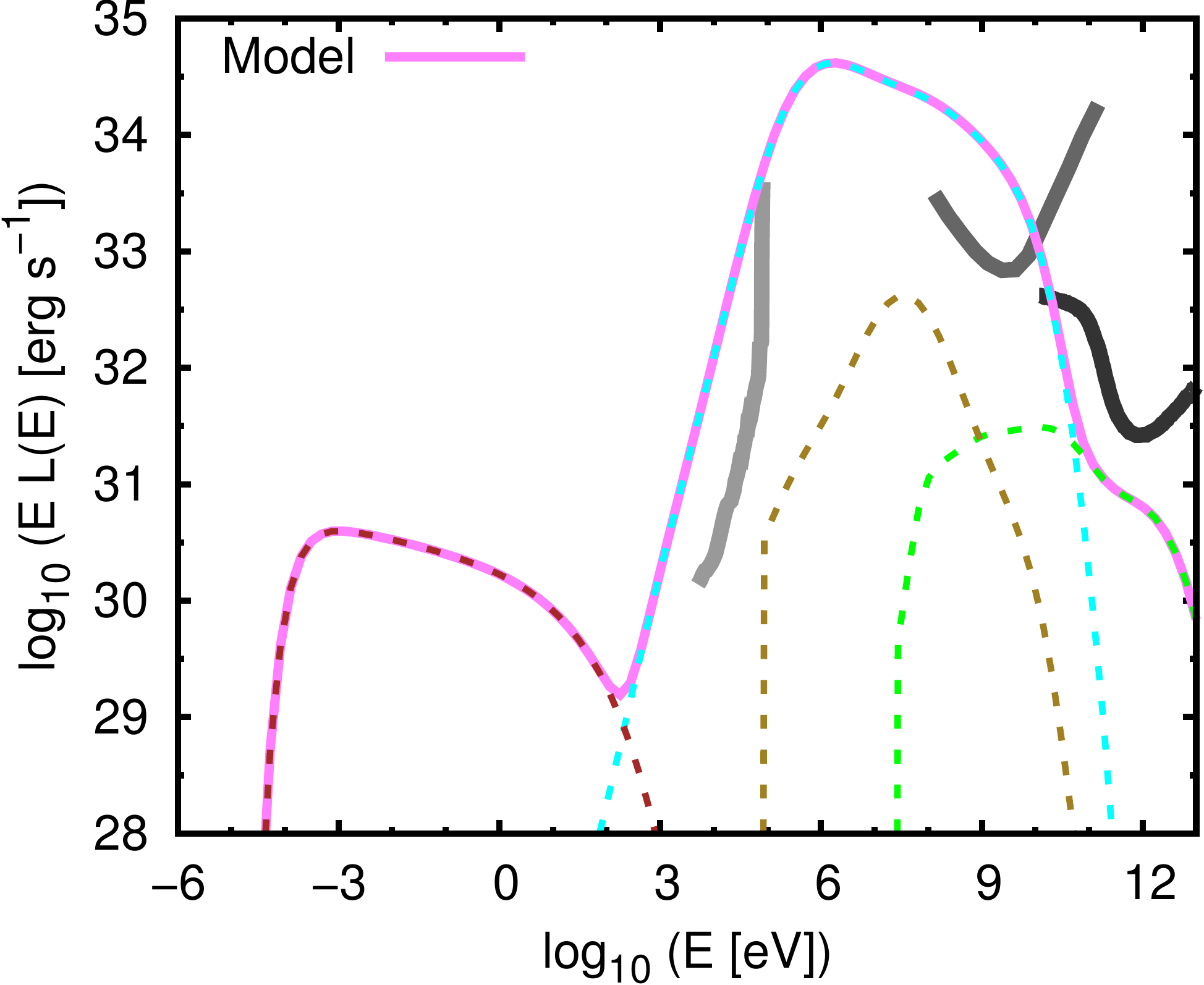}
    \caption[]{Broadband SEDs for the epoch of radio observations (\textit{top}), roughly the present epoch (\textit{middle}), 
    and roughly the periastron passage (\textit{bottom}). For the epoch of radio observations we show the ATCA data from 2008-2009, and 
    upper-limits
    in the X-ray flux derived from \cite{2006PASA...23...50B}. Dashed lines indicate contributions from synchrotron (brown), IC (cyan),
    relativistic Bremsstrahlung (olive), and \textit{p-p} (green); the filled red curve is the total emission. We show instrument 
    sensitivity curves for 1-Ms \textit{NuSTAR} (grey), 4-yr \textit{Fermi} (dark grey), and 50-h CTA (black).}
    \label{fig:seds_nohard}
\end{figure}

The broadband SEDs for different epochs are shown in Fig.~\ref{fig:seds_nohard}, along with some instrument sensitivities\footnote{The 
sensitivity curve of \textit{NuSTAR} is adapted from \cite{Koglin2005}, whereas the \textit{Fermi} one is from 
\texttt{http://fermi.gsfc.nasa.gov}, and the CTA one from \cite{Takahashi2013}.}. 

As the stars get closer, the density of the stellar photon field becomes higher. The IC cooling time decreases in consequence,
leading to an increase in the IC luminosity and a slight decrease in the electron maximum energy.
We predict that the system will be detectable in hard X-rays and HE $\gamma$-rays close to periastron passage, even if $f_\mathrm{NT}$
is an order of magnitude below the assumed 0.2 value. The production 
of TeV photons will be low and undetectable with the current or near-future instruments.

\subsubsection{Synthetic maps} \label{sec:maps}

The full information from the spatially resolved radio observations is only partially contained in the SEDS. Therefore we compare 
the morphology predicted by our model with the observed one \citep{Ben2015}. For that purpose we produce synthetic 
radio maps at two frequencies, 2.3~GHz and 8.6~GHz, and for different epochs. To produce the radio maps we first project the 3D 
emitting structure in the plane of the sky, obtaining a two-dimensional distribution of flux at a certain frequency. Then we cover 
the plane adjusting at each location of the map an elliptic Gaussian 
that simulates the synthesized (clean) beam. If the observational synthesized beam has an angular size $a\times b$, each 
Gaussian has $\sigma_x=a/\sqrt{8\log2}$ and $\sigma_y=b/\sqrt{8\log2}$.
At each pointing we sum the emission from every location weighted by the distance between its projected position and the Gaussian center. 

The LBA observations at 2.3~GHz from \cite{Ben2015} had a synthesized beam of $15\times 11$~mas$^2$; at 8.6~GHz, we adopt a 
synthesized beam of $\sim 4\times 3$~mas$^2$. 

\begin{figure}
    \centering
    \includegraphics[width=0.26\textwidth, angle=270]{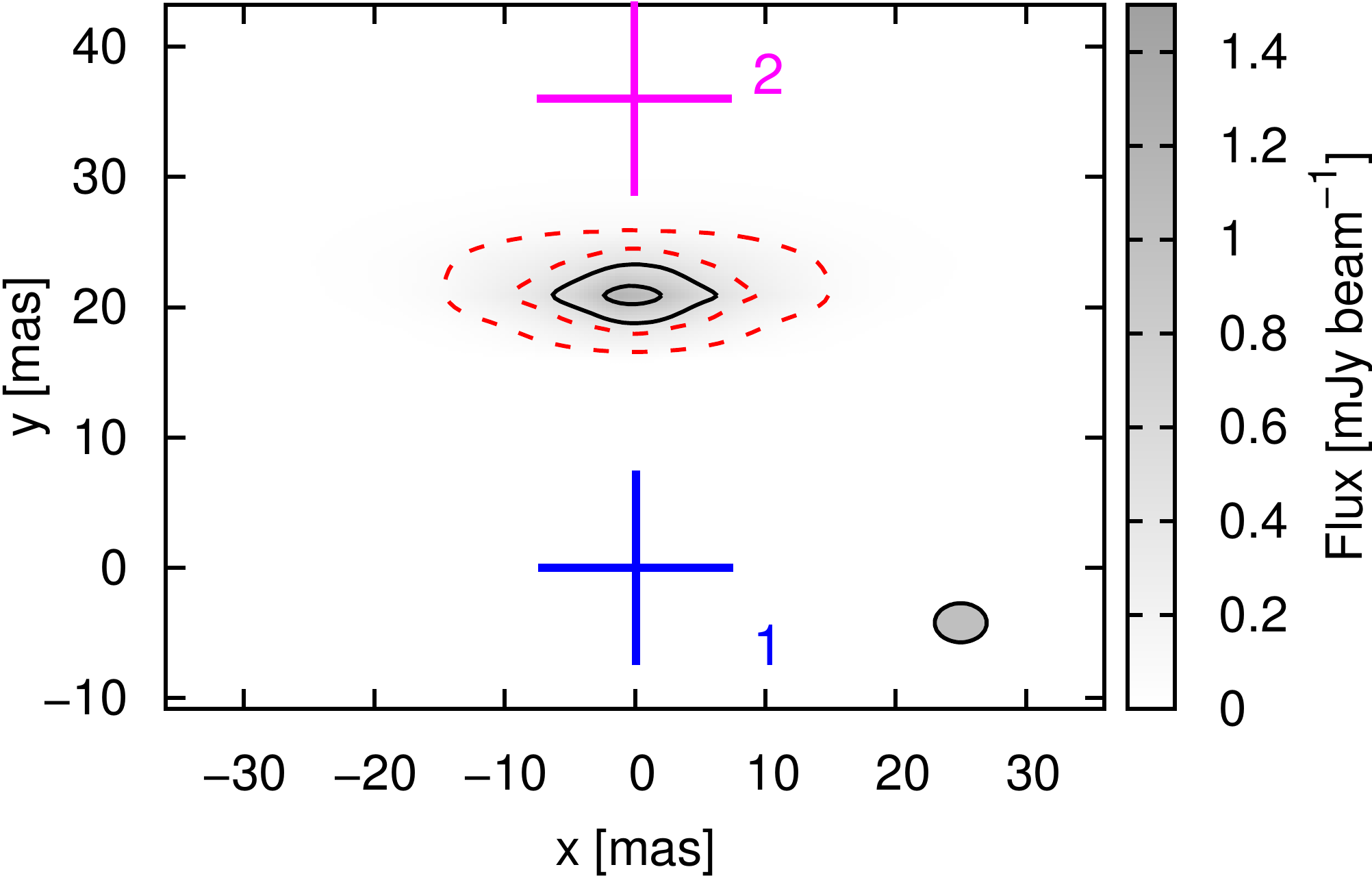}
    \includegraphics[width=0.26\textwidth, angle=270]{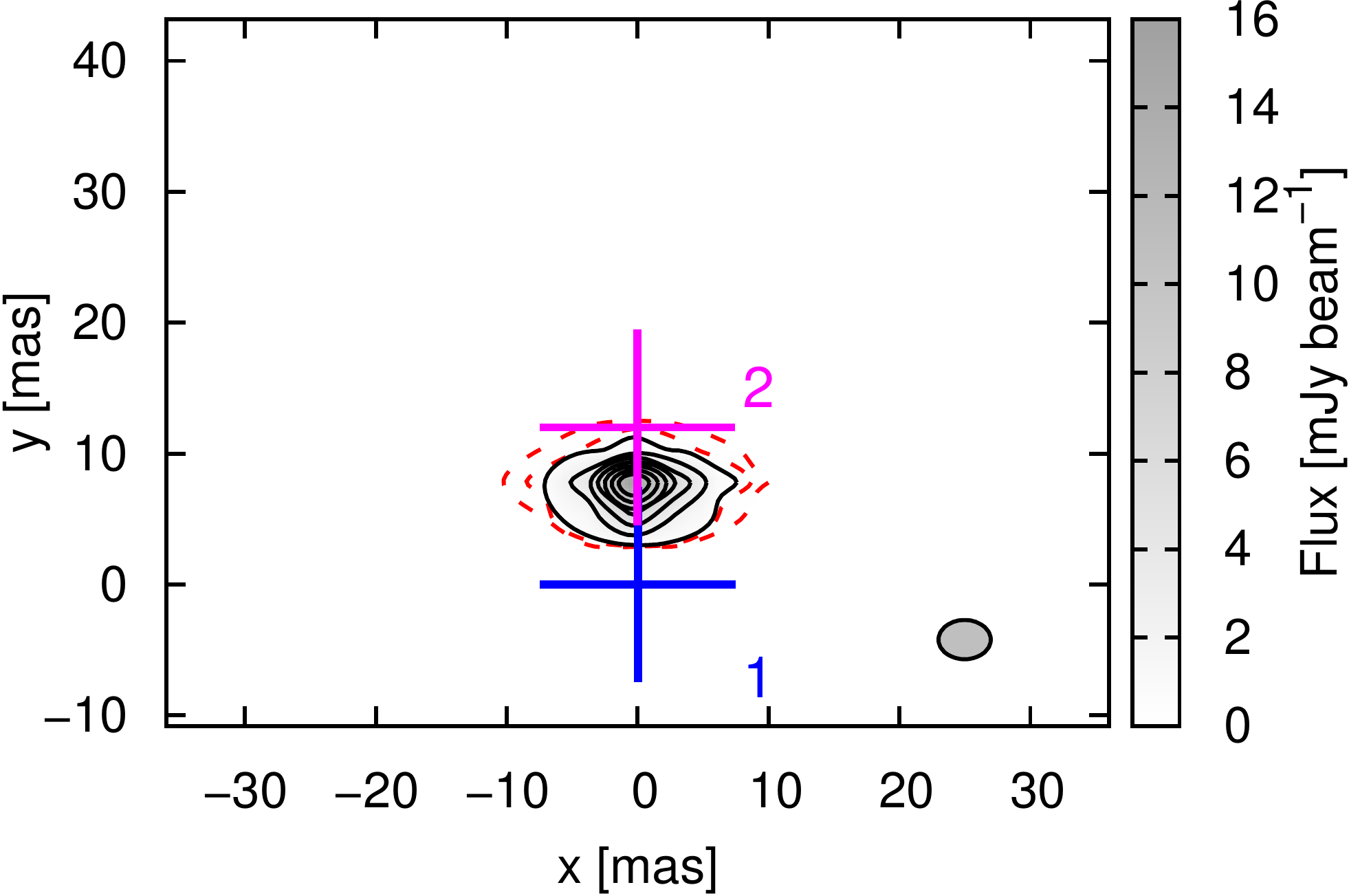}
    \includegraphics[width=0.26\textwidth, angle=270]{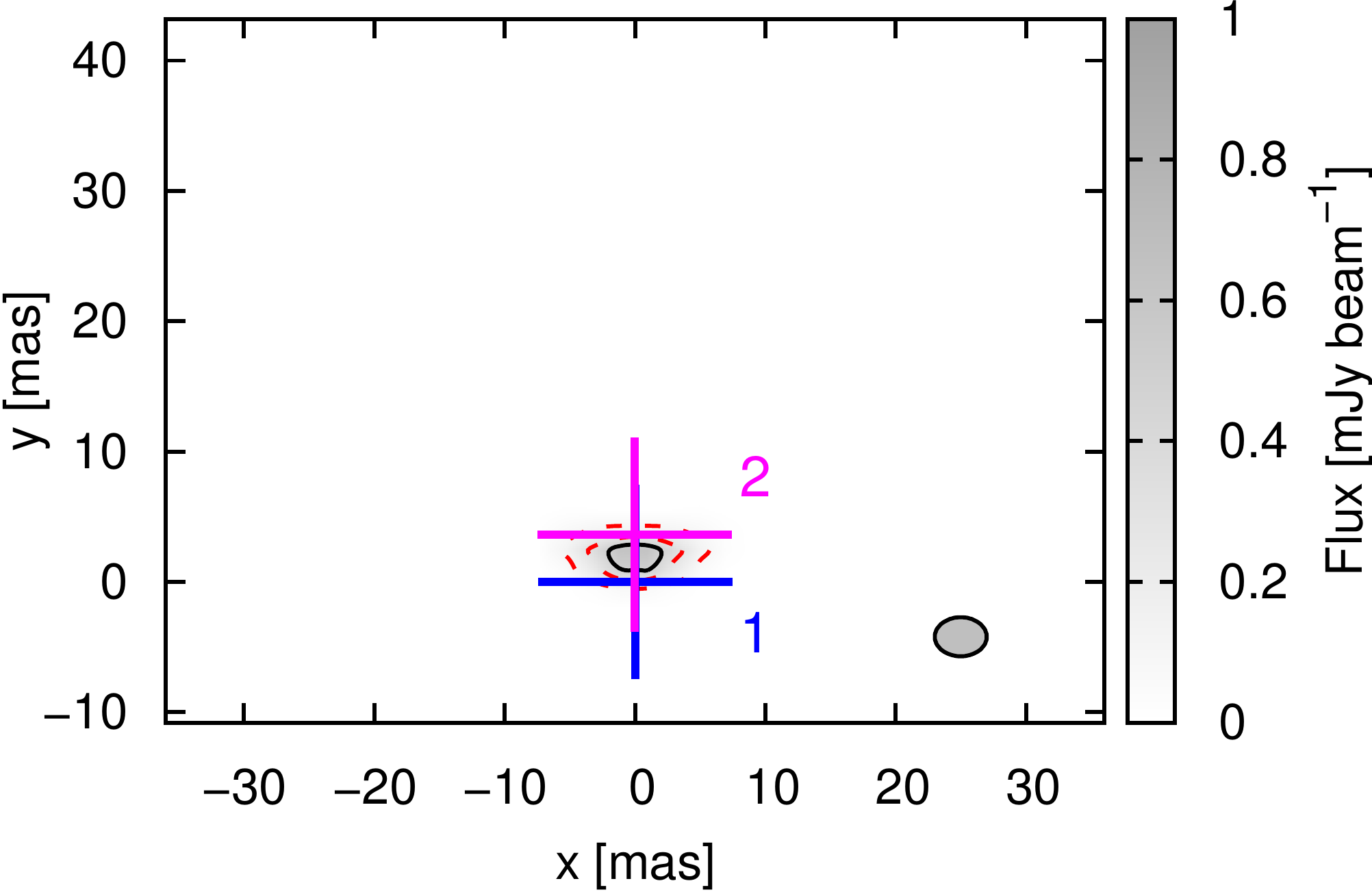}
    \caption[]{Synthetic radio maps at 8.6~GHz for the same binary separation as in the epoch of the LBA observations (\textit{top}), 
    for roughly the present epoch (\textit{middle}), and for roughly the periastron passage (\textit{bottom}) for the 
    model described in Sec.~\ref{subsec:lowB}. In all images the red contours are at 0.1 mJy beam$^{-1}$ and 0.3 mJy beam$^{-1}$. In 
    the \textit{top} image the black contours are 0.5 mJy beam$^{-1}$ and 1.0 mJy beam$^{-1}$, in the \textit{middle} image they start at 
    0.5 mJy beam$^{-1}$ and increase in 2.0 mJy beam$^{-1}$, and in the \textit{bottom} one there is only one contour at 0.5 mJy beam$^{-1}$.}
    \label{fig:maps8GHz}
\end{figure}

As seen in Fig.~\ref{fig:maps2GHz_variousi}, for a high orbital inclination the angular size of the WCR is much larger, and the resultant 
image is completely different from the one in the maps obtained from observations \citep{Ben2015}. This can be explained as follows: the 
projected distance between the stars is $D_{\mathrm{proj}} = 36$ mas, so for a source at 2.3 kpc the linear distance 
is $D \sim 83 \cos(i)^{-1}$~AU. A large ($>60^\circ$) value of $i$ leads to $D > 150$ AU. The impact of FFA is also 
very much reduced in this case. The linear size of the dominant WCR radio-emitting region is of a few $D$, and as 
the wind density drops as $r^{-2}$, the FFA is very small except for photons coming from a region close to the apex 
of the WCR. This results in a small region of the whole WCR being affected by absorption.
In contrast, Fig.~\ref{fig:maps2GHz_variousi} shows that for a low inclination ($i<30^\circ$) the morphology and flux levels in the 
synthetic map are consistent with the observations. The emission maps at 2.3~GHz (not shown) for future epochs simply indicate that the 
radiation will come from a much compact (unresolved) region, with a faint signal hardly detectable above the noise level.
Nevertheless, the maps at 8.6~GHz presented in Fig.~\ref{fig:maps8GHz} show that future 
observations at this frequency should be able to track the evolution of the WCR, as it is bright enough.

\subsection{Equipartition magnetic field scenario} \label{subsec:strongB}

We apply the same model as in Sect.~\ref{subsec:lowB}, but with a different value of $\zeta$ in order to analyze a scenario 
with an extremely high-magnetic field. We explore the case $\zeta = 1$, which corresponds to equipartition between the magnetic field and the 
thermal energy densities in the WCR (Eq.~5). Note that under such condition the magnetic field would be dynamically relevant, i.e., the 
magnetic pressure would be a significant fraction of the total pressure in the post-shock region. Nonetheless,
we do not alter our prescription of the flow properties, as we adopt a phenomenological model for 
them only to get a rough approximation of the gas properties in the shock. Moreover, our intentions are just to give a semi-qualitative
description of this extreme scenario, not to make a precise modeling of the emission, and to
obtain a rough estimate of some relevant physical parameters. Fixing $\zeta = 1$ yields $B \sim 1$~G in the 
WCR, and, if the magnetic field in the WCR is solely due to adiabatic compression of the stellar magnetic field lines, then we get
$B_{*1} \sim 1.3$~kG and $B_{*2} \sim 2.2$~kG on the stellar surface of each star. 
Note that such high values of $B_*$ have never been measured in PACWBs \citep{Neiner2015}, 
something that suggests that in this scenario the magnetic field in the WCR should be generated or amplified \textit{in situ}.
In this case, due to the stronger magnetic field, electrons with Lorentz factor 
$\gamma_\mathrm{e} \approx 20$ emit synchrotron radiation at $\nu \approx 1$~GHz, and in accordance the electron energy 
distribution cutoff is set to $\gamma_\mathrm{min} = 20$. Setting $f_\mathrm{NT} \approx 10^{-4}$ leads to a spectral fit of the 2009 observed radio 
fluxes with $\chi^2 = 2.1$, and the obtained synthetic radio maps are very similar to the ones obtained in Sect.~\ref{subsec:lowB}. However, 
such a small value of $f_\mathrm{NT}$ seems suspicious. The synchrotron
cooling time is $t_\mathrm{sy} \approx t_\mathrm{IC}$ for $E_\mathrm{e} < 10$ GeV, when IC interactions occur in the Thomson regime, 
and $t_\mathrm{sy} < t_\mathrm{IC}$ for $E_\mathrm{e} > 10$ GeV, when IC interactions occur in the Klein-Nishina regime. Therefore 
the bulk of the NT emission is produced in the form of low-energy photons, and little luminosity goes into $\gamma$-ray production. The value 
of $L_\gamma \lesssim 10^{31}$~erg s$^{-1}$ obtained is $\sim 10^4$ times smaller than the one obtained in Sect.~\ref{subsec:lowB}. 
Considering that we are using a value $10^4$ times larger for $\zeta$, this result is expected as it follows the relation 
$L_\mathrm{IC}/L_\mathrm{sy} \propto B^{-2} \propto \zeta^{-1}$ \citep[see, e.g.,][]{White1995}. In this scenario HD~93129A is not 
detectable with the current or in progress HE observatories.

 \begin{figure}
    \centering
    \includegraphics[width=0.29\textwidth, angle=270]{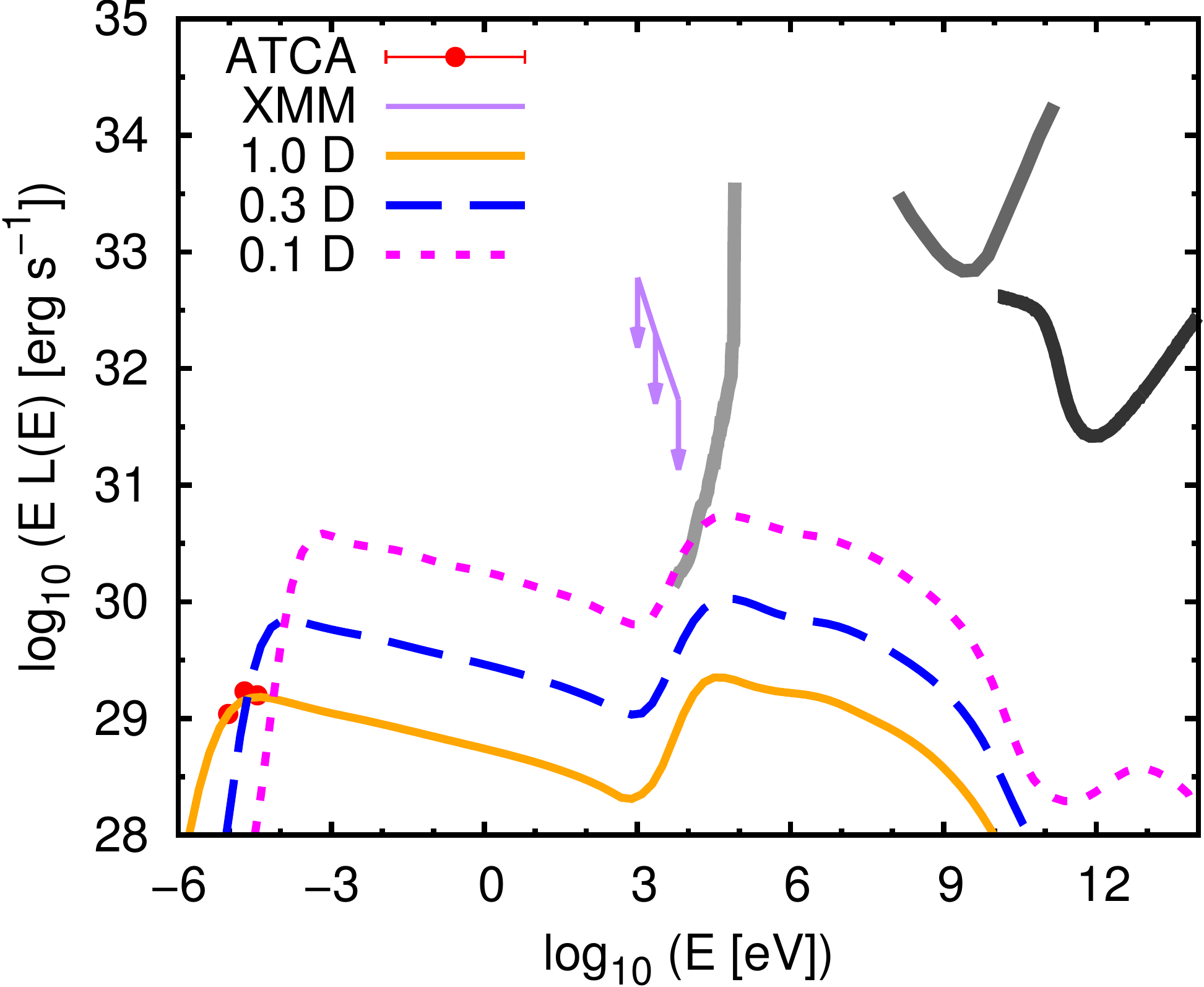}
    \caption[]{Same as Fig.~\ref{fig:seds_nohard} but considering an equipartition magnetic field in the WCR. The solid orange curve represents 
    the modeled SED for the epoch of radio observations, the long-dashed blue line is the SED for roughly the present epoch (\textit{middle}), 
    and the short-dashed magenta line is the SED for roughly the periastron passage.}
    \label{fig:seds_highB}
\end{figure}

\subsection{Best case scenario} \label{subsec:harden}

\begin{figure}
    \centering
    \includegraphics[width=0.29\textwidth, angle=270]{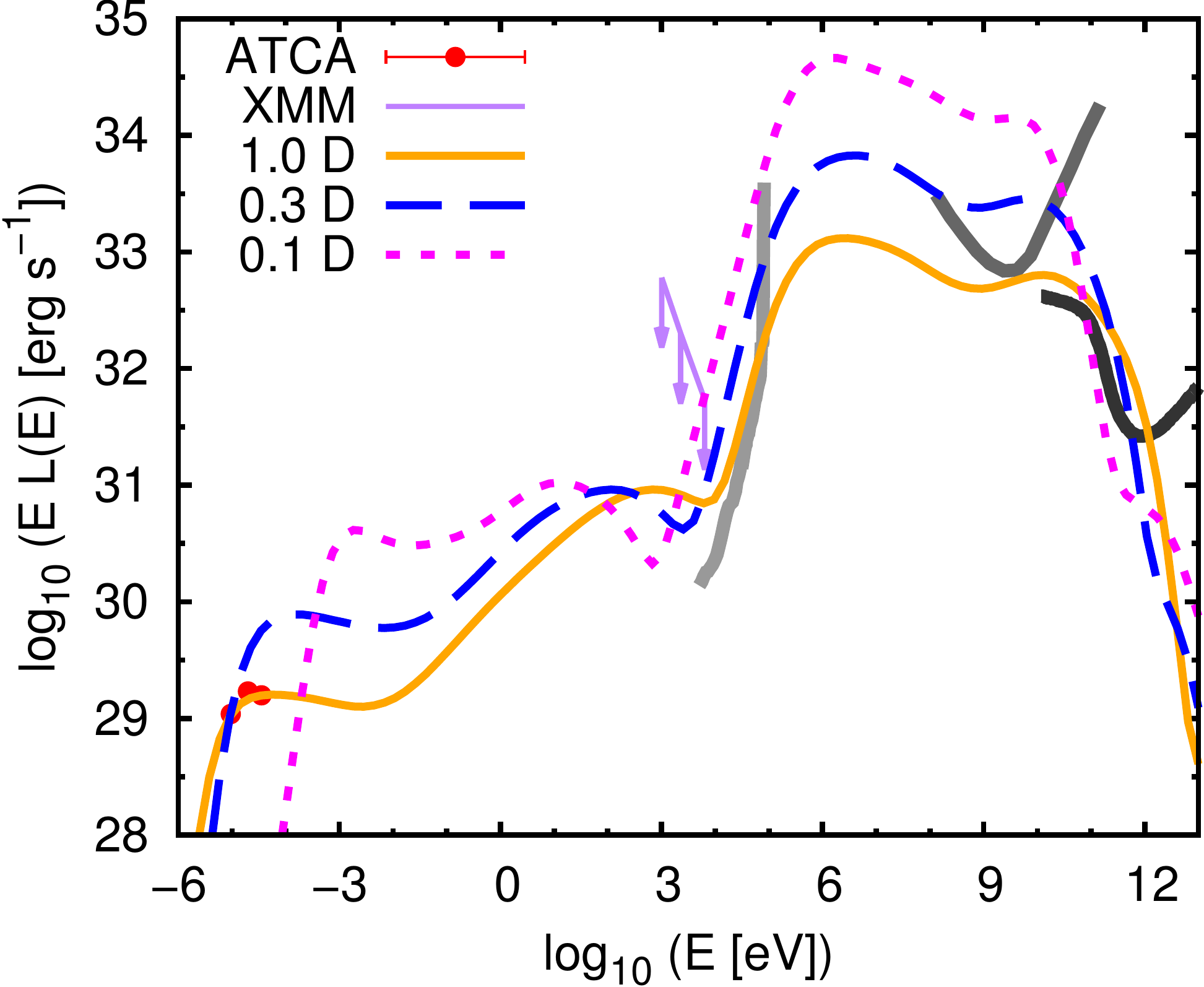}
    \caption[]{Same as Fig.~\ref{fig:seds_nohard} but considering a possible hardening in the electron distribution. The solid orange curve represents 
    the modeled SED for the epoch of radio observations, the long-dashed blue line is the SED for roughly the present epoch (\textit{middle}), 
    and the short-dashed magenta line is the SED for roughly the periastron passage.}
    \label{fig:seds_hard}
\end{figure}

In this case we fix the HE particle spectrum to optimize the NT outcome. We maximize the HE outcome by assuming 
essentially the same parameters as in the scenario presented in Sect.~\ref{subsec:lowB}, but also that electrons
with Lorentz factors $> 7\times 10^3$ are accelerated with $p = 2$. This may be a reasonable assumption, as discussed in 
Sect.~\ref{sec:distribution}. The emission at radio frequencies is therefore similar to the one in Sect.~\ref{subsec:lowB}; 
the major difference is the stronger IC emission of $\gamma$-rays, as seen in the SEDs presented in Fig.~\ref{fig:seds_hard}. 
At energies close to
10~GeV, the source would be well within the detection capability of \textit{Fermi} when the binary is close to periastron. 
At energies $> 100$ GeV the source should be detectable by CTA. Note that the specific value of the turnover in 
the particle energy distribution has been chosen just to illustrate a case in which the chances of high-energy detection are enhanced.

\subsection{Hadronic scenario} \label{subsec:hadronic}

The secondary $e^\pm$ pairs from \textit{p-p} interactions could account for a significant part of the radio emission \citep[e.g.][]{Orellana2007}.
To explore this possibility, we consider the most favorable scenario in which the totality of the wind kinetic energy injected 
in the WCR goes into accelerating NT particles (i.e., $f_{\rm NT}=1$), from which almost 100\% goes into NT protons. For this NT 
proton energetics and under the density values of the shocked wind, the $\gamma$-ray luminosity from $pp$ interactions is 
$L_{pp}\sim 3\times10^{30}-7\times10^{31}$~erg s$^{-1}$, for $p\sim 3.2-2$ for protons (the higher luminosity corresponds to a higher $\zeta$).
Taking into account that the luminosity from $pp$ interactions to 
secondary $e^\pm$ pairs is $\kappa \, L_{pp}$, with $\kappa \approx 0.25-0.5$ depending on $p$ \citep{Kelner2006},
the synchrotron luminosity of the secondary pairs in the radio band can be estimated as:
\begin{equation}
L_\mathrm{sec,rad} \sim \kappa \, L_{pp} \frac{t_\mathrm{esc}}{t_\mathrm{sy,rad}}\,.
\end{equation}
The secondary $e^\pm$ energy distribution at creation will be similar to that of the primary protons, and given that it should have a cutoff around 
$10-50$ MeV to explain the radio data, a cutoff in the proton energy distribution should be also introduced. 
For the escape time one can derive $t_\mathrm{esc} = D_*/v_\mathrm{esc} \sim 3\times10^6$~s, where $D_*$ is the shock distance to the 
closest star, and $v_\mathrm{esc}$ is the typical shocked wind velocity. The synchrotron cooling time of the secondary $e^\pm$ 
pairs depends on their energy and the local magnetic field. For the pairs that produce synchrotron emission at the observed 
frequencies, we get $t_\mathrm{sy} \sim 10^{10}$~s if $B \sim 0.01\,B_\mathrm{eq}$, 
and $t_\mathrm{sy} \sim 10^{6}$~s for $B \sim B_\mathrm{eq}$. Therefore, if the magnetic field is below
$0.1\,B_\mathrm{eq}$, we get $L_\mathrm{sec,sy} < 10^{29}$ erg s$^{-1} = L_\mathrm{sy,obs}$, which is too low to account for the observed 
luminosity. If, instead, the magnetic field is above $0.1\,B_\mathrm{eq}$, then $L_\mathrm{sec,sy} \gtrsim L_\mathrm{sy,obs}$. However, 
such a strong magnetic field leads to an excessively high synchrotron luminosity of the primary 
electrons unless only a tiny amount $<10^{-3}$ ($<10^{-4}$ if $B \sim B_\mathrm{eq}$) of the injected NT luminosity goes into these particles. 

We have shown qualitatively that even if we adopt an extreme case of $f_\mathrm{NT}=1$, which is probably far from realistic, 
we cannot satisfactorily explain the observed 
radio emission with a hadronic scenario unless the $B\gtrsim 0.1\,B_\mathrm{eq}$ and for the primary electrons $f_\mathrm{NT_e}< 10^{-3}$ in 
the WCR. Nevertheless, a small contribution of secondary $e^\pm$ pairs to radio emission cannot be discarded.


\section{Conclusions}

 In this work we studied in detail the information provided by the observations of the colliding-wind binary HD~93129A. We developed a 
 leptonic model that reproduces fairly well the radio observations, and we showed that it is possible to explain the low-energy cutoff 
 in the SED by means of a combination of FFA in the stellar winds and a low-energy cutoff in the electron energy distribution. We also 
 provide radio morphological arguments to favour a low $i$ for the system, implying a high eccentricity but also a not far-in-future
 periastron passage. Unfortunately, the limited knowledge on the physical parameters of the source, some model degeneracy in the fraction of the 
 energy that goes into accelerating non-thermal particles versus the magnetic field intensity in the wind-collision region, 
 and limited observations, make it difficult to fully constrain the properties of the non-thermal processes in HD~93129A. In any case, the 
 results presented in this work provide a good insight for future observational campaigns in the radio and $\gamma$-ray range. In particular,
 we showed that future observations of the system HD~93129A at $0.5-10$ GHz radio frequencies can confirm the nature of the radio 
 absorption/suppression mechanism by comparing them with our predictions of the cutoff frequency and the steepness of the turnover.
 Observations in the hard X-ray range ($10-100$ keV), moreover, would provide tighter constraints to the free parameters 
 in our model, such as the injected particle energy distribution at low energies, 
 and the magnetic field strength, through comparison of the synchrotron and IC emission, constraining then the acceleration 
 efficiency. We have shown that moderate values of the stellar surface magnetic field ($B_*<50$ G) are sufficient to account for the synchrotron
 emission even if there is no amplification of the magnetic field besides adiabatic compression; however, if the magnetic field in the 
 wind-collision region is high (as it would be suggested if the source remains undetected in hard X-rays and $\gamma$-rays), then 
 magnetic field amplification is likely to occur. 
 We predict that this source is a good candidate to be detected during the periastron passage by \textit{Fermi} in HE, 
 and even possibly by CTA in VHE, although it is not expected to be a strong and persistent $\gamma$-ray emitter.
   
 The electron energy distribution spectral index is poorly constrained, specially at high energies. However, even for a soft index $p = 3.2$ 
 the $\gamma$-ray emission could still be detected by \textit{Fermi}. Considering that slightly smaller values of $p \approx 3$, are also
 derived from some radio observations, and that a possible hardening of the particle energy distribution cannot be discarded, the detection
 of HD~93129A in $\gamma$-rays in the near future seems feasible as long as $f_\mathrm{NT} \gtrsim 10^{-2}$ and $\zeta \lesssim 10^{-3}$. 
 Given the very long period of the binary HD~93129A, we strongly suggest to dedicate observing time to this source in $\gamma$-rays, 
 but also in the radio and X-ray band, during the next few years, as this will provide very valuable information of non-thermal processes 
 in massive colliding-wind binaries. 
     

\begin{acknowledgements}
This work is supported by ANPCyT (PICT 2012-00878).
V.B-R. acknowledges financial support from MICINN and European Social Funds through a Ram\'on y Cajal fellowship.
V.B-R. acknowledges support by the MDM-2014-0369 of ICCUB (Unidad de Excelencia 'Mar\'ia de Maeztu').
This research has been supported by the Marie Curie Career Integration Grant 321520.
V.B-R. and G.E.R. acknowledges support by the Spanish
Ministerio de Econom\'{\i}a y Competitividad (MINECO)
under grant AYA2013-47447-C3-1-P. The acknowledgment extends to the whole GARRA group.
\end{acknowledgements}

\bibliographystyle{aa} 
\bibliography{biblio} 

\end{document}